\documentclass[preprint,aps]{revtex4-2}
\usepackage{mathrsfs}
\usepackage{graphicx}
\usepackage{multirow}
\usepackage{makecell}
\usepackage{booktabs}
\usepackage{color}
\usepackage{bm}

\usepackage{hyperref}
\hypersetup{
    colorlinks,
    citecolor=black,
    filecolor=black,
    linkcolor=black,
    urlcolor=black
}
\begin{document}

\title{Hidden Moir\'e Topology of Low-Symmetry Weyl Surfaces}

\author{Cong Li$^{1,\sharp,*}$, Zhilong Yang$^{2,\sharp,*}$, Hongxiong Liu$^{3}$, Magnus H. Berntsen$^{1}$, Francesco Scali$^{4}$, Dibya Phuyal$^{1}$, Jianfeng Zhang$^{3}$, Timur K. Kim$^{5}$, Jacek Osiecki$^{6}$, Balasubramanian Thiagarajan$^{6}$, Youguo Shi$^{3}$, Tao Xiang$^{3}$, Quansheng Wu$^{3}$, Oscar Tjernberg$^{1,*}$
}

\affiliation{
\\$^{1}$Department of Applied Physics, KTH Royal Institute of Technology, Stockholm 11419, Sweden
\\$^{2}$School of Mathematics and Physics, University of Science and Technology Beijing, Beijing 100083, China
\\$^{3}$Beijing National Laboratory for Condensed Matter Physics, Institute of Physics, Chinese Academy of Sciences, Beijing 100190, China
\\$^{4}$Dipartimento di Fisica, Politecnico di Milano, Piazza Leonardo da Vinci 32, 20133 Milano, Italy
\\$^{5}$Diamond Light Source, Harwell Campus, Didcot, OX11 0DE, United Kingdom
\\$^{6}$MAX IV Laboratory, Lund University, 22100 Lund, Sweden
\\$^{\sharp}$These people contributed equally to the present work.
\\$^{*}$Corresponding authors: conli@kth.se, zhlyang@ustb.edu.cn, oscar@kth.se
}


\maketitle

%%Abstract

\begin{center}
{\bf Abstract}
\end{center}

{\bf Topological materials are defined by the correspondence between bulk topology and boundary states, yet this correspondence becomes enigmatic on low-symmetry surfaces where bulk and surface periodicities are inherently mismatched. Here we reveal a hidden moir\'e topology emerging on the (103) surface of the Weyl semimetal NdAlSi. Angle-resolved photoemission spectroscopy uncovers closed Fermi-arc loops and momentum-space moir\'e modulations, phenomena unanticipated in conventional topological theory. We show that these emerge from incomplete bulk projection and multi-cell interference governed by a least-common-multiple framework. Least-common-multiple guided DFT and Green's-function calculations quantitatively reproduce the observed spectra, establishing the universality of this commensuration rule. These findings transform a long-standing paradox of bulk-boundary correspondence into a new paradigm of momentum-space moir\'e reconstruction, bridging crystalline and quasicrystalline topologies and opening routes to flat-band engineering on complex surfaces.\\}

%%Introduction
\noindent {\bf Introduction}

Since the discovery of topological materials, a central question in condensed-matter physics has been how bulk topology manifests at material boundaries. Nowhere is this connection more striking than in Weyl semimetals\cite{KManna_NRM2018_CFelser,NPArmitage_RMP2018_AVishwanath,BQLv_RMP2021_HDing,XGWan_PRB2011_SYSavrasov,SYXu_Science2015b_MZHasan,BQLv_PRX2015_HDing,LXYang_NP2015_YLChen,HMWeng_PRX2015_XDai,SMHuang_NC2015_MZHasan,KHashimoto_PTEP2017_XWu,KGomi_PTEP2022}, where open Fermi arcs on high-symmetry surfaces such as (001)/(100) of tetragonal or orthorhombic lattices\cite{SYXu_Science2015b_MZHasan,BQLv_PRX2015_HDing,LXYang_NP2015_YLChen,CLi_NC2023_OTjernberg,CLi_AM2025_OTjernberg,CLi_PNAS2025_OTjernberg} and the (0001) facet of hexagonal lattices\cite{YWang_PRB2022_XJZhou,CLi_CP2025_JVDBrink} directly connect the surface projections of bulk Weyl points. These canonical surfaces, where bulk and surface periodicities are trivially commensurate, have accordingly dominated both experimental and theoretical studies of topological semimetals.

In contrast, low-symmetry surfaces remain largely unexplored\cite{ZJXu_AM2013_MHXie,BSingh_PRB2016_ABansil,XHZheng_NSR2021_RRDu}, despite offering a qualitatively new regime of topology. Their intrinsic incommensurability between bulk and surface Brillouin zones (SBZs) disrupts the conventional one-to-one mapping between bulk Weyl points and surface Fermi arcs, raising the question of how bulk-boundary correspondence (BBC) manifests when translational commensurability is lost. Whether BBC in this regime survives in a modified form or requires a new organizing principle remains unknown, as no direct experimental observation or coherent theoretical framework has yet addressed it. This gap stems from formidable challenges on both fronts: experimentally, low-symmetry facets rarely cleave into atomically uniform planes\cite{XHZheng_NSR2021_RRDu}, and even when achieved, disorder and multiple terminations obscure intrinsic surface states in angle-resolved photoemission spectroscopy (ARPES) measurements; theoretically, conventional slab models assume commensurate periodicities and therefore fail to capture the correct surface repeat unit, making density functional theory (DFT) reproduction of experimental spectra notoriously difficult. Because Weyl semimetals themselves originate from broken symmetries, low-symmetry surfaces naturally extend this principle into the boundary, where reduced crystalline constraints are expected to generate qualitatively new boundary phenomena--ranging from anisotropic or reconstructed Fermi arcs\cite{CLi_AM2025_OTjernberg,ZRao_Nature2019_HDing,DSSanchez_Nature2019_MZHasan} to nonlinear transport\cite{ISodemann_PRL2015_LFu,FdeJuan_NC2017_JEMoore,YTokura_NC2018_NNagaosa,NNagaosa_ARCMP2024_YYanase} and higher-order boundary states\cite{LTrifunovic_PRX2019_PWBrouwer}. A framework capable of reconciling bulk-surface incommensurability--both conceptually and computationally--is therefore essential for advancing this frontier.

Here, we overcome these barriers by realizing a well-defined (103) surface of the Weyl semimetal NdAlSi through precise pre-orientation and pre-cutting of single crystals, followed by controlled cleavage that exposes atomically uniform terraces. This preparation enables high-resolution ARPES measurements that reproducibly resolve multiple surface-state configurations. The spectra reveal an apparent mismatch between bulk-projected Weyl points and the SBZ--a direct manifestation of bulk-surface incommensurability. Guided by this observation, we show that the mismatch is not a breakdown of BBC but a natural outcome of incomplete bulk projection. By extending bulk-to-surface mapping beyond the first BZ, we establish a least common multiple (LCM) criterion that restores bulk-surface commensuration and explains the emergence of a ``new smaller SBZ'', corresponding to the momentum-space moir\'e pattern observed experimentally. For NdAlSi (103), surface-projected DFT calculations based on this framework quantitatively reproduce the ARPES spectra and capture the characteristic interference among inequivalent surface cells. These results demonstrate that the observed moir\'e modulation is the direct consequence of LCM-guided reconstruction, where finite phase offsets between neighboring surface cells generate long-period interference and hybridization between surface and bulk-projected states. More broadly, this universal framework resolves the apparent BBC paradox and establishes low-symmetry surfaces as momentum-space analogues of moir\'e systems, linking incommensurability, surface reconstruction, and correlated boundary phenomena across a broad class of quantum materials.\\

\noindent {\bf Results}
%Figure1

\begin{figure*}[tbp]
\begin{center}
\includegraphics[width=1\columnwidth,angle=0]{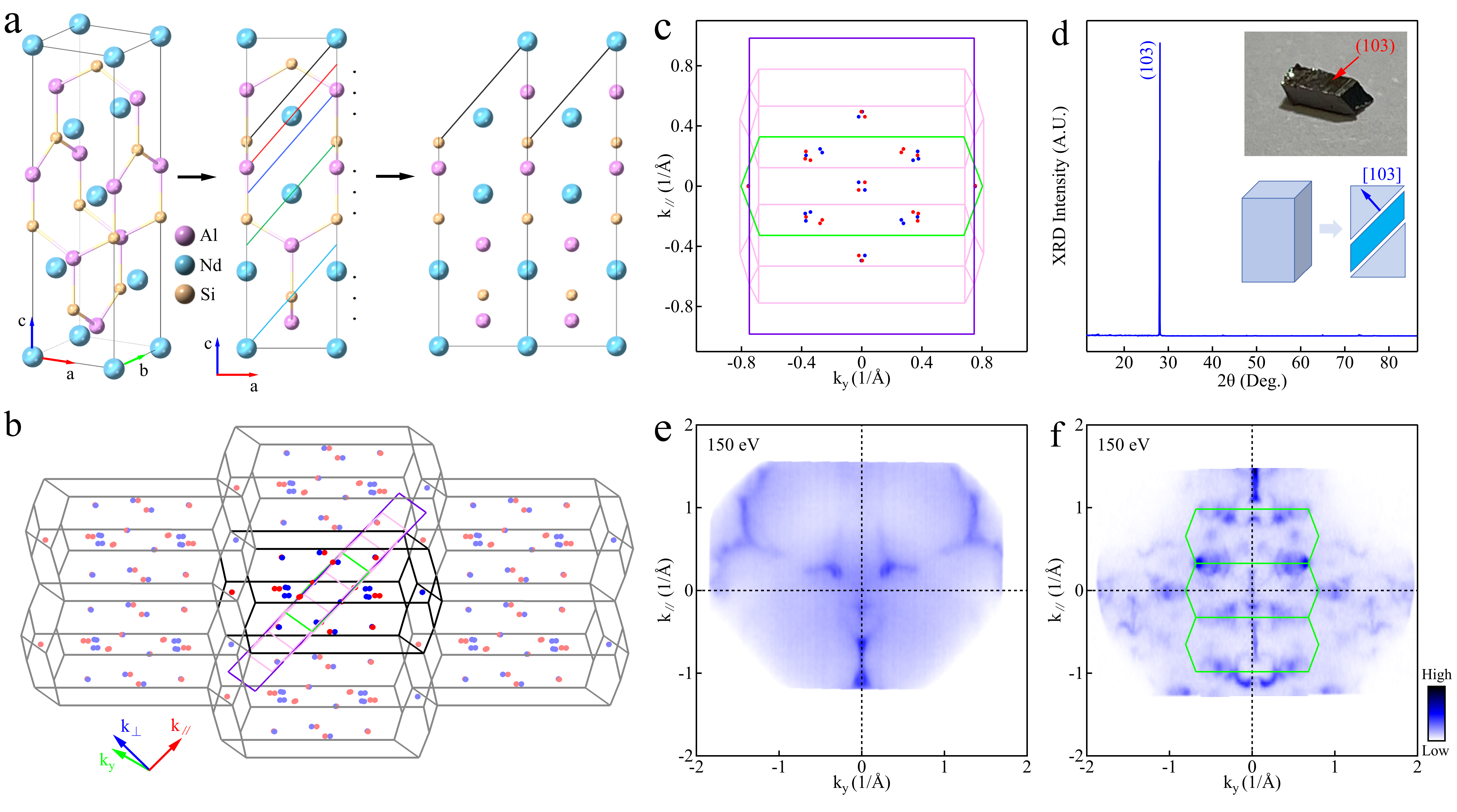}
\end{center}
\caption{\footnotesize\textbf{Crystal structure and electronic projections of NdAlSi (103) surface.} (a) Crystal structure of NdAlSi with the (103) surface cut. The differently colored cuts represent the potential cleavage planes. The rightmost panel shows one possible cleavage plane. (b) BZ construction and projection onto the (103) surface. The red (blue) dots corresponding to Weyl points with chirality $+1$ ($-1$). (c) Periodicity mismatch between the SBZ and the first bulk BZ projection. The red (blue) dots corresponding to Weyl points with chirality $+1$ ($-1$). (d) XRD confirming the exposed (103) surface. (e, f) Fermi surfaces measured at 150 eV on uneven (disordered) (e) and flat (well defined) (f) regions of the (103) surface. Purple lines denote the (103) theoretical SBZ, whose size is determined by the (103) cross section of half a primitive unit cell. Since the body-centered lattice halves the real-space periodicity along the [001] direction, the corresponding BZ along $k_{z}$ is doubled. For consistency, the (103) SBZ is likewise defined by the cross section of half a primitive unit cell (see Section 4 of the Supplementary Materials for the details). Green lines denote the experimentally observed periodicity extracted from (f); pink lines denote the first bulk BZ projected onto the (103) surface.
}
\label{1}
\end{figure*}

We begin with the Weyl semimetal NdAlSi\cite{CLi_NC2023_OTjernberg,CLi_AM2025_OTjernberg,CLi_PNAS2025_OTjernberg,JGaudet_NM2021_CLBroholm}, a non-centrosymmetric member of the RAlSi (R = rare-earth) family\cite{RLuo_PRB2023_SBorisenko,APSakhya_PRM2023_MNeupane,YCZhang_CP2023_MYi,HTRong_CM2025_CYChen}, which hosts well-characterized Weyl points and high-quality single crystals ideally suited for surface-sensitive spectroscopy. We focus on the low-symmetry (103) surface, which has never been explored yet provides a natural setting to probe how BBC manifests when bulk and surface periodicities are incompatible. The crystallographic structure and orientation of the (103) surface are shown in Fig.~\ref{1}a, while Fig.~\ref{1}b illustrates the bulk BZ and its projection. Unlike high-symmetry facets such as (001) or (100), the (103) projection exhibits an intrinsic mismatch between the periodicities of the SBZ (purple lines) and the projected bulk BZ (pink lines), as shown in Fig.~\ref{1}c--this incommensurability forms the experimental starting point of our study.

The intrinsic incommensurability revealed by geometric projection motivates direct experimental verification on an atomically defined (103) surface. The surface orientation was confirmed by X-ray diffraction (XRD), which verifies the high-quality exposure of the (103) plane (Fig.~\ref{1}d; see Methods for sample preparation). Fermi-surface maps acquired at 150 eV further highlight the crucial role of surface quality: uneven regions (Fig.~\ref{1}e) show diffuse features dominated by bulk states, whereas well-defined terraces (Fig.~\ref{1}f) exhibit sharp surface states that follow the periodicity of the surface Brillouin zone (SBZ, purple lines). A closer inspection, however, reveals an additional modulation with a smaller period (green lines), indicating a more subtle reconstruction. On the uneven region, the suppression of coherent surface states\cite{CLi_PNAS2025_OTjernberg} exposes bulk-like bands whose periodicity does not coincide with that of the SBZ. These observations demonstrate that the mismatch between bulk and surface periodicities is not merely a geometric construct but a directly observable phenomenon on the (103) surface. This raises a central question: if Weyl points follow the bulk BZ periodicity while surface Fermi arcs adhere to the SBZ, does the BBC break down on low-symmetry surfaces?

%Figure2

\begin{figure*}[tbp]
\begin{center}
\includegraphics[width=1\columnwidth,angle=0]{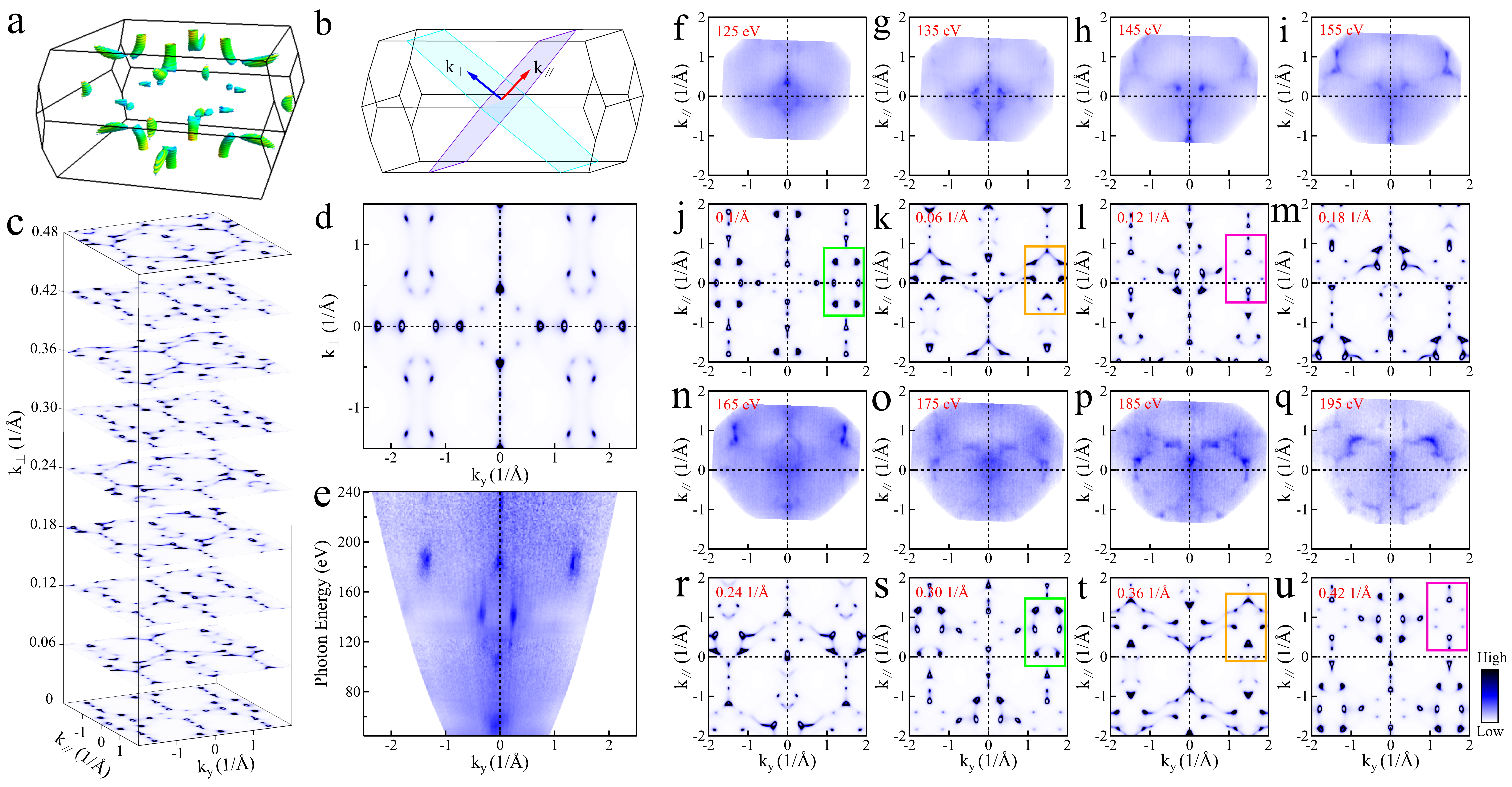}
\end{center}
\caption{\footnotesize\textbf{Bulk electronic structures of the (103) surface.} (a) DFT calculated 3D Fermi surface of NdAlSi. (b) The (103) surface and its perpendicular cross-section in momentum space are expected to be probed with photon energies ranging from approximately 125 eV to 205 eV. (c) DFT calculated bulk Fermi surface of the momentum-space slices along the [103] direction ($k_{\perp}$). (d) DFT-calculated bulk Fermi surface of the plane [cyan plane in (b)] perpendicular to the (103) surface [purple plane in (b)]. (e) Photon-energy dependent ARPES spectral intensity map at the Fermi level along the $k_{y}$ direction on a relatively strong disordered (103) surface, where both surface states and SBPSs are nearly suppressed. (f-i, n-q) Photon energy dependent bulk Fermi surface mappings (125-195 eV) revealing $k_{\perp}$ dispersion.  (j-m, r-u) Corresponding DFT calculations. Each slice produces a shifted replica of the bulk Fermi surface within the SBZ, reflecting the intrinsic periodicity of bulk states along the [103] direction; the green, orange, and pink boxes highlight features that are shifted as $k_{\perp}$ changes. The features do not appear strictly identical here because only discrete $k$-slices are shown, but denser sampling would reveal exact repetitions.
}
\label{2}
\end{figure*}

To understand the nature of the electronic states on the (103) surface and answer the above question, we performed systematic ARPES measurements combined with bulk DFT calculations. The 3D bulk Fermi surface from DFT is shown in Fig.~\ref{2}a, while Fig.~\ref{2}b illustrates the (103) surface orientation and its perpendicular cross-section in momentum space. Calculated bulk Fermi surface slices along the [103] direction (k$_{\perp}$) and perpendicular to the (103) surface (cyan plane in Fig.~\ref{2}b) are presented in Figs.~\ref{2}c and~\ref{2}d, serving as theoretical benchmarks.

Experimentally, the photon energy dependent ARPES spectral intensity map at the Fermi level along k$_{y}$ (Fig.~\ref{2}e) reveals strong k$_{\perp}$ modulation, a hallmark of bulk-derived states. Systematic Fermi surface measurements over photon energies from 125 eV to 195 eV (Figs.~\ref{2}f-i,~\ref{2}n-q) further confirm this k$_{\perp}$ dispersion. The corresponding DFT calculations (Figs.~\ref{2}j-m,~\ref{2}r-u) reproduce the experimental features with remarkable accuracy. More detailed bulk Fermi surfaces of the NdAlSi (103) surface are provided in Section 3 of the Supplementary Materials.

Beyond establishing the bulk band structure, these results provide critical insight into the apparent breakdown of the BBC. In particular, the evolution of bulk Fermi surface slices along [103] reveals a systematic in-plane translation of features as k$_{\perp}$ is varied. Each slice produces a shifted replica of the bulk Fermi surface within the SBZ, reflecting the intrinsic periodicity of bulk states along the [103] direction. As successive bulk BZs are considered, these replicas accumulate into an effective superlattice periodicity that is distinct from the primitive bulk BZ but naturally commensurate with the SBZ.

From this perspective, the apparent mismatch between the periodicity of projected Weyl points and that of the surface Fermi arcs is a direct consequence of restricting the projection to the first bulk BZ. The bulk projections along [103] demonstrate experimentally that multiple BZs must be included: only then does the accumulated superlattice restore the periodicity needed to match the surface states. This insight provides the experimental foundation for resolving the BBC paradox, as developed below.

%Figure3

\begin{figure*}[tbp]
\begin{center}
\includegraphics[width=0.9\columnwidth,angle=0]{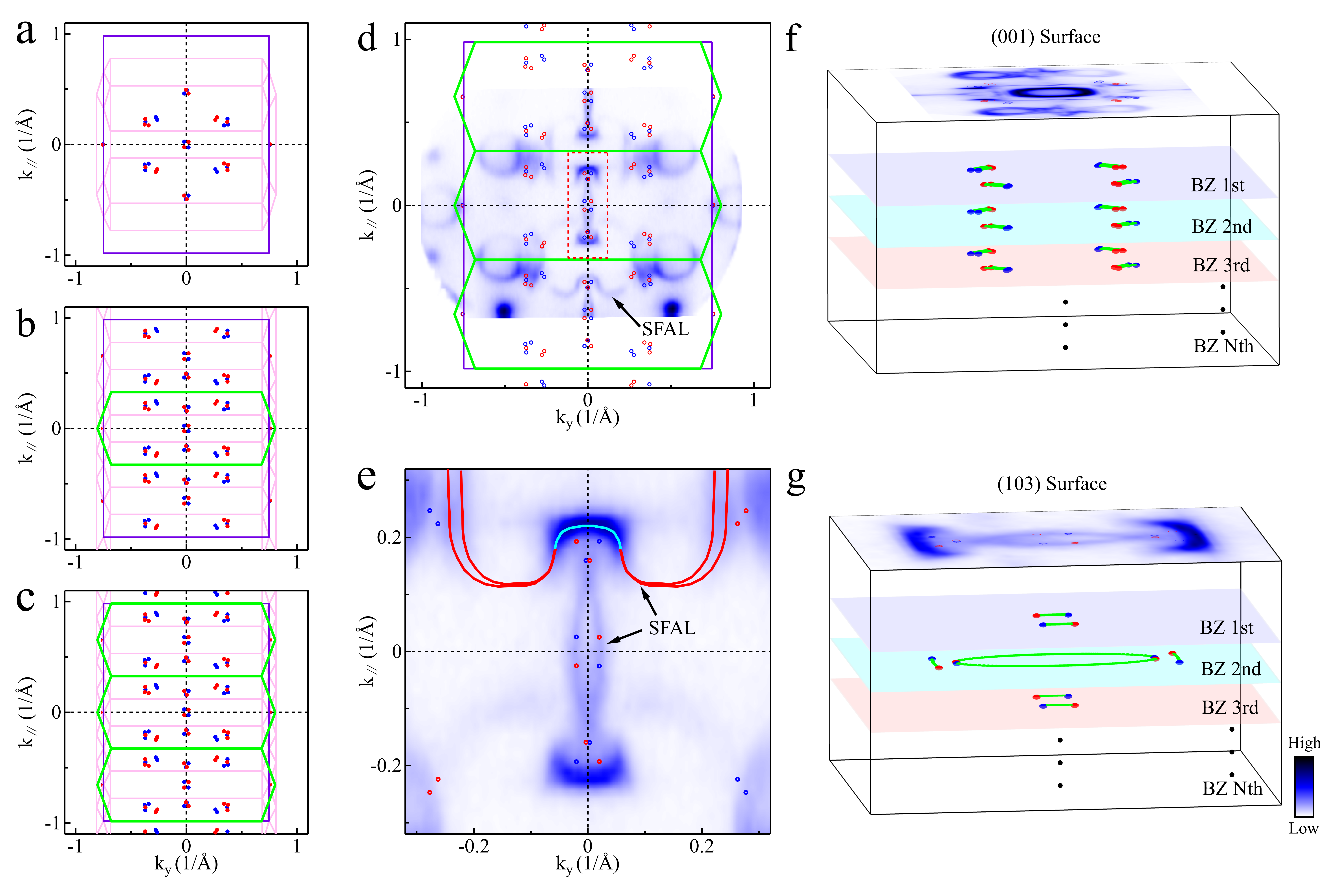}
\end{center}
\caption{\footnotesize\textbf{Resolving the BBC paradox on the (103) surface.} (a) Projection of Weyl points onto the (103) surface considering only the first bulk BZ, showing a mismatch with the surface periodicity. (b-c) Inclusion of the second-order (b) and third-order (c) bulk BZs along $k_{z}$ progressively restores the periodicity and brings the bulk projection into agreement with the surface states. Red (blue) dots denote Weyl points of chirality $+1$ ($-1$). (d-e) Comparison between the bulk projection Weyl points and the measured surface states demonstrates that multiple BZs are required to recover the correct periodicity. (e) is the magnified view of (d). The red and cyan lines represent surface Fermi arcs, which hybridize to form SFALs as indicated by the black arrows. (f-g) General scheme for different surfaces: for the (001) surface (f), the first bulk BZ is sufficient, whereas for the (103) surface (g), more bulk BZ projections are need to generate a superlattice periodicity consistent with the SBZ. The Fermi surface in (g) is an enlarged view of the red dashed boxed region in (d). Purple lines denote the (103) theoretical SB; green lines denote the newly formed (103) SBZ, corresponding to the SBZ observed in experiment; pink lines denote the first bulk BZ projected onto the (103) surface. The connection of the surface Fermi arcs in panel (g) is not the actual connectivity, but rather a schematic illustration indicating that Fermi arcs from different bulk BZ may hybridize upon projection onto the (103) surface (see Section 6 of the Supplementary Materials for details). 
}
\label{3}
\end{figure*}

To clarify how this resolves the apparent BCC paradox, we track how the bulk-projection periodicity evolves as higher-order BZs are included. If only the first bulk BZ is considered, the projected Weyl points (Fig.~\ref{3}a) clearly mismatch the theoretical SBZ (purple lines) and the experimental SBZ (green lines). Adding the second-order (Fig.~\ref{3}b) and third-order (Fig.~\ref{3}c) BZs and Weyl points along $k_{z}$ progressively restores the periodicity, bringing the bulk projection closer to the surface states. This restoration follows a LCM criterion: the projected displacement between successive bulk BZs closes only after an integer multiple of the in-plane reciprocal vectors. For the (103) surface of NdAlSi, the minimum closure occurs after $N_{\min}=3$ steps (see Section 4 in the Supplementary Materials). As illustrated in Fig.~\ref{3}d, the projected Weyl points recover the theoretical SBZ periodicity (purple lines) once higher-order BZs are included. 

Beyond this restoration, the higher‐order projections reveal an emergent smaller repeating unit: a “new SBZ” (green lines in Fig.~\ref{3}d) whose period along $k_{\parallel}$ is one third of the theoretical SBZ. This “new SBZ” coincides with the one observed experimentally. Equivalently, this effect can be viewed as a fixed lateral shift between successive bulk projections (see Section 4 in the Supplementary Materials), which manifests as a momentum-space moir\'e pattern--a weak superlattice modulation with a period one third of the original SBZ (Fig.~\ref{3}d). In general, whether a single bulk BZ suffices depends on surface symmetry. For high-symmetry surfaces such as (001) of NdAlSi, the first bulk BZ already coincides with the SBZ (Fig.~\ref{3}f). For low-symmetry surfaces such as (103), however, successive bulk projections are laterally shifted (Fig.~\ref{3}g), and only after $N_{\min}$ steps does the bulk periodicity close. Exact spectral reconstruction is achieved only in the infinite-projection limit (see Section 4 in the Supplementary Materials). Thus, the observed mismatch is not a breakdown of the BBC but rather the natural consequence of incomplete bulk projection, while the emergent new SBZ reflects the moir\'e-like interference among multiple shifted projections. Here we display one representative surface-state configuration; additional terminations observed on the same cleaved surface are presented in Section 5 of the Supplementary Materials. 

An additional subtlety arises when considering the surface Fermi arcs themselves. On high-symmetry surfaces such as (001) of NdAlSi, arcs from successive bulk BZs project in perfect registry, typically preserving the canonical picture of open arcs directly connecting projected Weyl points (Fig.~\ref{3}f). In sharp contrast, on low-symmetry surfaces like (103), the incommensurate periodicities of bulk and SBZs cause consecutive arcs to project with lateral mismatches. This misalignment naturally drives hybridization between surface arcs originating from bulk zones that are displaced both in-plane and out-of-plane (Fig.~\ref{3}g), so that the observed features are not simple replicas but collective superpositions. In certain cases, multiple arcs merge into closed SFALs--a boundary topology unusual on high-symmetry facets (Fig.~\ref{3}d; see Section 6 of the Supplementary Materials for details). 

Our framework, together with the best ARPES resolution presently achievable, provides compelling evidence for the emergence of SFALs. These loop-like states enrich the boundary phenomenology of Weyl semimetals: their closed $k$-space trajectories provide natural channels for quantum interference and oscillatory responses, while the concentrated Berry curvature they enclose may enhance nonlinear or nonreciprocal transport. Moreover, overlapping arc segments can promote nesting instabilities, offering a route to correlated boundary orders. Future high-resolution probes such as spin-resolved ARPES or quantum oscillations will provide complementary confirmation and deeper insight into the microscopic structure of these loops.

%Figure4

\begin{figure*}[tbp]
\begin{center}
\includegraphics[width=1\columnwidth,angle=0]{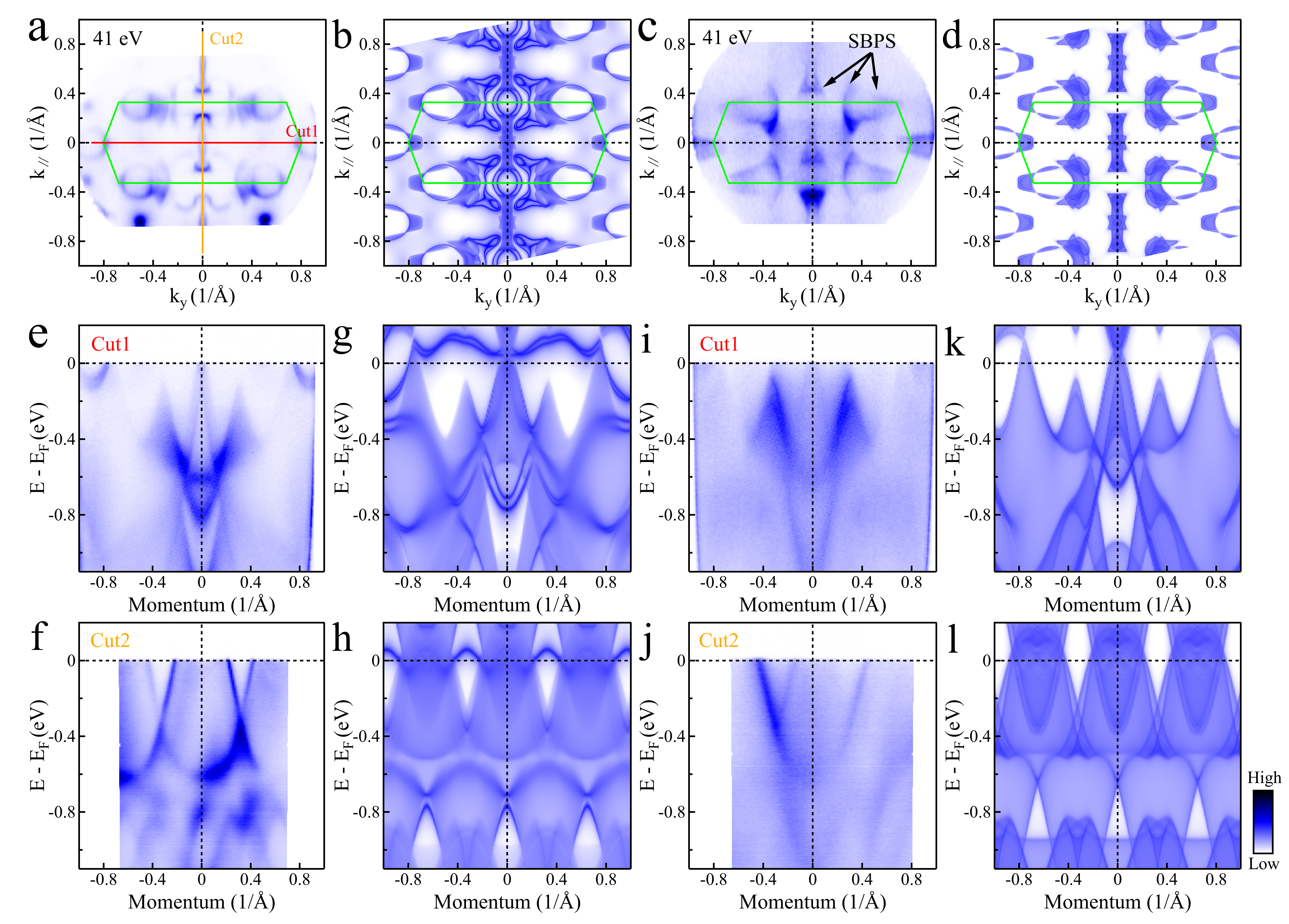}
\end{center}
\caption{\footnotesize\textbf{Comparison of experimental and DFT results for surface states and SBPSs on NdAlSi (103) surface.} (a) Fermi surface of the NdAlSi (103) surface measured on a flat region with the photon energy of 41 eV, revealing surface states. (b) Surface projected DFT calculation of the (103) Fermi surface for a Si atom terminated surface cleavage at the Si-Al layer (see Section 5 in the Supplementary Materials for details). (c) Impurity-induced disorder filters out surface states\cite{CLi_PNAS2025_OTjernberg}, exposing SBPS. (d) DFT calculation of the corresponding SBPSs on the (103) surface. Green lines indicate the experimental SBZ, consistent with Fig.~\ref{1}e. (e-f) Band dispersions measured along momentum Cut1 [red line in (a), (e)] and Cut2 [orange line in (a), (f)]. (g-h) Surface projected DFT band structures for the Si-terminated (103) surface along momentum Cut1 (g) and Cut2 (h). (i-j) Band dispersions along the same cuts measured on the disordered surface, where comparison with calculations (k-l) confirms the assignment of surface states and SBPSs.
}
\label{4}
\end{figure*}

To capture such arc reconstruction and hybridization theoretically, the surface periodicity used in modeling must be commensurate with bulk projections. This requirement is fulfilled by the LCM criterion, which provides a systematic prescription for selecting appropriate surface unit cells in DFT slab calculations to ensure commensurability between bulk projections and surface periodicities. In the case of the (103) surface of NdAlSi, the minimal commensuration requires $N_{\min}=3$, corresponding to a three-cell surface supercell composed of three consecutive unit cells. By contrast, a single-cell slab (or restricting to the first bulk BZ) systematically fails to reproduce the experimental periodicity, because the projection closure remains incomplete. With the correct three-cell supercell, surface-projected DFT captures the reduced periodicity and dominant features of the ARPES spectra. For a Si-terminated (103) surface (Fig.~\ref{4}b), the LCM-consistent calculation reproduces the emergent “new smaller SBZ”, originating from the phase offsets of $2\pi/3$ between bulk BZs of three consecutive, nonequivalent (103) surface unit cells. Their hybridization yields the reduced periodicity observed in experiment. 

Despite this overall agreement, finer aspects of the Fermi-arc connectivity are not fully reproduced in the idealized calculations. In ARPES, closed surface Fermi-arc loops (SFALs) appear within the reduced SBZ, whereas the LCM-consistent DFT yields only open arcs. This discrepancy most likely reflects two limitations of the model: first, the absence of zigzag-type reconstructions or mixed local terminations, which in real samples enhance coupling between surface and bulk states; and second, the incomplete treatment of hybridization across multiple projected bulk BZs, whose lateral mismatch further promotes loop-like connectivity. We therefore regard the observed SFAL as an emergent property of low-symmetry surfaces beyond what is captured by ideal slab calculations. In addition to the Si-terminated regions, ARPES measurements also reveal two further surface-state configurations on other areas of the same cleaved (103) surface (see Fig. S10 in the Supplementary Materials). These distinct spectra share the reduced periodicity but do not match any of the terminations tested in Fig. S11 of the Supplementary Materials, suggesting a more complex origin such as zigzag-type reconstructions or mixed local terminations beyond the simple three-cell picture in Fig. S9b in the Supplementary Materials. A detailed discussion of these additional states, and their relation to the LCM criterion, is provided in Section 5 of the Supplementary Materials.

Beyond the ordered regions, disorder offers a complementary perspective on the same framework. When spectra are collected on a disordered area of the (103) surface, sharp surface states are strongly suppressed (Fig.~\ref{4}c)\cite{CLi_PNAS2025_OTjernberg}, effectively acting as a natural “filter” that allows otherwise hidden surface bulk-projected states (SBPSs) to emerge. Remarkably, these SBPSs are well reproduced by the same surface-projected DFT calculations (Fig.~\ref{4}b), underscoring the robustness of the LCM-guided modeling. When a surface state enters the bulk-projected region, it hybridizes with the bulk continuum and evolves into a surface bulk-resonant state (SBRS)\cite{CSeibel_PRL2015_HEbert,JSBarriga_PRB2016_QRader,NEhlen_PRB2018_AGruneis,MGuttler_NC2019_DVVyalikh,JSchusser_PRL2022_FReinert,QLu_QF2022_DQian,WJLiu_PRB2022_SQiao,XXZhang_PNAS2024_NNagaosa}, producing hybridized features observable in ARPES. Momentum-resolved band dispersions along two representative cuts (Cut1 and Cut2) are shown in Figs.~\ref{4}e and~\ref{4}f. These sharp dispersive bands agree well with surface-projected DFT calculations (Figs.~\ref{4}g and~\ref{4}h). By contrast, along the same cuts on the disordered surface (Figs.~\ref{4}i and~\ref{4}j), broader and bulk-like bands appear, consistent with calculated SBPSs (Figs.~\ref{4}k and~\ref{4}l). Section 8 of the Supplementary Materials provide more details on the identification of surface states, bulk states, SBPSs, and SBRSs.\\

%Summary

\noindent {\bf Discussion and conclusion}

By investigating the low-symmetry (103) surface of the Weyl semimetal NdAlSi, we uncover a previously hidden regime of topological boundary reconstruction that fundamentally extends the conventional notion of BBC. While Weyl points follow the intrinsic bulk periodicity, the surface Fermi arcs adhere to a reduced periodicity defined by the SBZ. What initially appears as a paradox of the BBC is in fact a consequence of incomplete bulk projection, where successive bulk BZs generate laterally phase-shifted replicas that close only after multiple unit cells. This reconstruction restores commensuration through a LCM criterion, naturally producing the emergent “new smaller SBZ” observed in ARPES. The resulting interference pattern manifests as a momentum-space moir\'e modulation, in which neighboring surface unit cells contribute discrete phase offsets analogous to twisted or displaced layers in real-space moir\'e systems.

This momentum-space moir\'e analogy carries profound implications. The periodic phase interference between inequivalent surface unit cells can compress surface dispersions and generate flat-band-like electronic structures, thereby amplifying correlation effects at the boundary. In this picture, low-symmetry surfaces emerge as natural moir\'e analogues in momentum space, where the interplay of topology, symmetry, and interference gives rise to boundary states beyond the canonical open-arc paradigm. The observation of surface Fermi-arc loops (SFALs) further demonstrates that low-symmetry surfaces host boundary topologies inaccessible on conventional facets, pointing to a new class of topological states governed by multi-cell commensuration and hybridization.

More broadly, the LCM framework provides a unified prescription for reconciling bulk and surface periodicities across complex crystals. Its mathematical structure generalizes naturally to Dirac and nodal-line systems, higher-order topological phases, and non-centrosymmetric superconductors. Beyond crystalline materials, the same principle extends to quasicrystals and aperiodic systems, where commensuration becomes infinite and quasi-periodicity replaces translational order. From a practical standpoint, the LCM criterion offers a robust guideline for determining the minimal surface supercell required for accurate DFT slab calculations--long a bottleneck in modeling low-symmetry surfaces.

In summary, our work transforms a perceived inconsistency of the BBC into a new paradigm of boundary reconstruction, establishing a bridge between topological surface physics and moir\'e interference phenomena. By revealing that incommensurability itself can serve as a design principle for emergent boundary states, these findings open pathways toward flat-band formation, enhanced correlations, and topological interference effects in both crystalline and quasiperiodic systems.\\

%Methods

\noindent {\bf Methods}\\
\noindent{\bf Sample growth} Single crystals of NdAlSi were grown from Al as flux. Nd, Al, Si elements were sealed in an alumina crucible with the molar ratio of 1 : 10 : 1. The crucible was finally sealed in a highly evacuated quartz tube. The tube was heated up to 1273 K, maintained for 12 hours and then cooled down to 973 K at a rate of 3 K per hour. Single crystals were separated from the flux by centrifuging. The Al flux attached to the single crystals were removed by dilute NaOH solution. 

\noindent{\bf Sample preparation of the (103) surface} To obtain the low-symmetry (103) surface of NdAlSi, we first determined the crystallographic orientation of the single crystals by single crystal XRD. Based on the identified orientation, the crystals were cut with a high-precision diamond wire saw to expose the desired (103) direction. Prior to mounting, a shallow laser pre-cut was introduced along the target orientation, which facilitated controlled cleavage of the sample inside the ARPES chamber. This procedure enabled reliable preparation of well-defined (103) surfaces suitable for spectroscopic measurements.

\noindent{\bf ARPES measurements} High-resolution ARPES measurements were performed at the I05 beamline of the Diamond synchrotron and at the Bloch beamline of the MAX IV synchrotron. The total energy resolution (analyzer and beamline) was set at 15$\sim$20 meV for the measurements. The angular resolution of the analyser was $\sim$0.1 degree. The beamline spot size on the sample was about 50 $\mu$m$\times$50 $\mu$m at the I05 beamline of the Diamond synchrotron and about 10 $\mu$m$\times$12 $\mu$m at the Bloch beamline of the MAX IV synchrotron. The samples were cleaved {\it in situ} and measured in ultrahigh vacuum with a base pressure better than 1.0$\times$10$^{-10}$ mbar at about 8 K at the I05 beamline of the Diamond synchrotron and about 18 K at the Bloch beamline of the MAX IV synchrotron.

\noindent{\bf DFT calculations} The electronic structure calculations for NdAlSi were performed based on the density functional theory (DFT)\cite{PHohenberg_PR1964_WKohn,WKohn_PR1965_LJSham} as implemented in the VASP package\cite{GKresse_CMS1996_JFurthmuller,GKresse_PRB1996_JFurthmuller}. The exchange-correlation functional was treated within the generalized gradient approximation (GGA) in the form of Perdew–Burke–Ernzerhof (PBE)\cite{JPPerdew_PRL1996_MErnzerhof}. The projector augmented wave (PAW) method\cite{PEBlochl_PRB1994,GKresse_PRB1998_DJoubert} was used to describe the electron–ion interactions. Since the measurement temperature is above the Curie temperature, NdAlSi is non-magnetic; therefore, the Nd pseudopotential without 4f electrons was employed in the calculations. A plane-wave kinetic energy cutoff of 360 eV was used. Brillouin zone sampling was performed using a 6×6×6 Monkhorst–Pack k-point grid \cite{HJMonkhorst_PRB1976_JDPack}. A Gaussian smearing width of 0.1 eV was applied to approximate the Fermi–Dirac distribution. Spin–orbit coupling (SOC) was included as a second-order perturbation. 

The bulk Fermi surface of NdAlSi was computed using a 60×60×60 k-point grid including SOC. The projection of bulk states onto the plane for different $k$-space slices was calculated via the Green's function method, using a Fermi broadening of 0.01 eV to obtain the spectral function.

For the (103) surface states and Fermi arcs, we employed an iterative Green's function method to model a semi-infinite surface and compute its surface states. Among the four possible terminations of the (103) surface, the Si-Al plane termination yielded results consistent with experimental observations.

\noindent {\bf Data Availability}

\noindent The authors declare that all data supporting the findings of this study are available within the paper and its Supplementary Information files.

\vspace{3mm}

\noindent {\bf Acknowledgement}\\
The authors thank Maxim Breitkreiz for his useful discussion of bulk-boundary correspondence. The work presented here was financially supported by the Swedish Research council (2019-00701 and 2019-03486) and the Knut and Alice Wallenberg foundation (2018.0104). Y.G.S. acknowledges the National Natural Science Foundation of China (Grants No. U2032204), the Informatization Plan of Chinese Academy of Sciences (CAS-WX2021SF-0102) and the National Key R\&D Program of China 2024YFA1408400. Q. S. W. acknowledges the National Natural Science Foundation of China No. 12274436. We acknowledge MAX IV Laboratory for time on Beamline BLOCH under Proposal 20241395 and 20250284 as well as Diamond Light Source for time on Beamline I05 under Proposal SI39652. Research conducted at MAX IV, a Swedish national user facility, is supported by the Swedish Research council under contract 2018-07152, the Swedish Governmental Agency for Innovation Systems under contract 2018-04969, and Formas under contract 2019-02496.

\vspace{3mm}

\noindent {\bf Author Contributions}\\
C.L. proposed and conceived the project. C.L. carried out the ARPES experiments with the assistance from M.H.B. and F.S.. Z.L.Y. contributed to the band structure calculations and theoretical discussions. H.X.L. and Y.G.S. contributed to NdAlSi crystal growth. C.L. and D.P. contributed to XRD measurements. C.L. developed the data analysis software, performed the data analysis, contributed to the theoretical derivation and wrote the paper. T.K., J.O. and B.T. provided the beamline support. C.L., Z.L.Y., M.H.B., D.P., J.F.Z., T.X., Q.S.W. and O.T. participate in the scientific discussions. O.T. revised the manuscript. All authors participated in and commented on the paper.

\noindent {\bf Competing Interests}\\
The authors declare no competing interests.

\newpage

\end{document}

\centerline{\large{Supplementary Material for}}
%\begin{center}
%\Large Supplementary Material for
%\end{center}

\title{Hidden Moir\'e Topology of Low-Symmetry Weyl Surfaces}

\author{Cong Li$^{1,\sharp,*}$, Zhilong Yang$^{2,\sharp,*}$, Hongxiong Liu$^{3}$, Magnus H. Berntsen$^{1}$, Francesco Scali$^{4}$, Dibya Phuyal$^{1}$, Jianfeng Zhang$^{3}$, Timur K. Kim$^{5}$, Jacek Osiecki$^{6}$, Balasubramanian Thiagarajan$^{6}$, Youguo Shi$^{3}$, Tao Xiang$^{3}$, Quansheng Wu$^{3}$, Oscar Tjernberg$^{1,*}$
}

\affiliation{
\\$^{1}$Department of Applied Physics, KTH Royal Institute of Technology, Stockholm 11419, Sweden
\\$^{2}$School of Mathematics and Physics, University of Science and Technology Beijing, Beijing 100083, China
\\$^{3}$Beijing National Laboratory for Condensed Matter Physics, Institute of Physics, Chinese Academy of Sciences, Beijing 100190, China
\\$^{4}$Dipartimento di Fisica, Politecnico di Milano, Piazza Leonardo da Vinci 32, 20133 Milano, Italy
\\$^{5}$Diamond Light Source, Harwell Campus, Didcot, OX11 0DE, United Kingdom
\\$^{6}$MAX IV Laboratory, Lund University, 22100 Lund, Sweden
\\$^{\sharp}$These people contributed equally to the present work.
\\$^{*}$Corresponding authors: conli@kth.se, zhlyang@ustb.edu.cn, oscar@kth.se
}

\pacs{}

\maketitle

{\bf 1. Distinguishing surface states from bulk states}

\begin{figure*}[tbp]
\begin{center}
\includegraphics[width=1\columnwidth,angle=0]{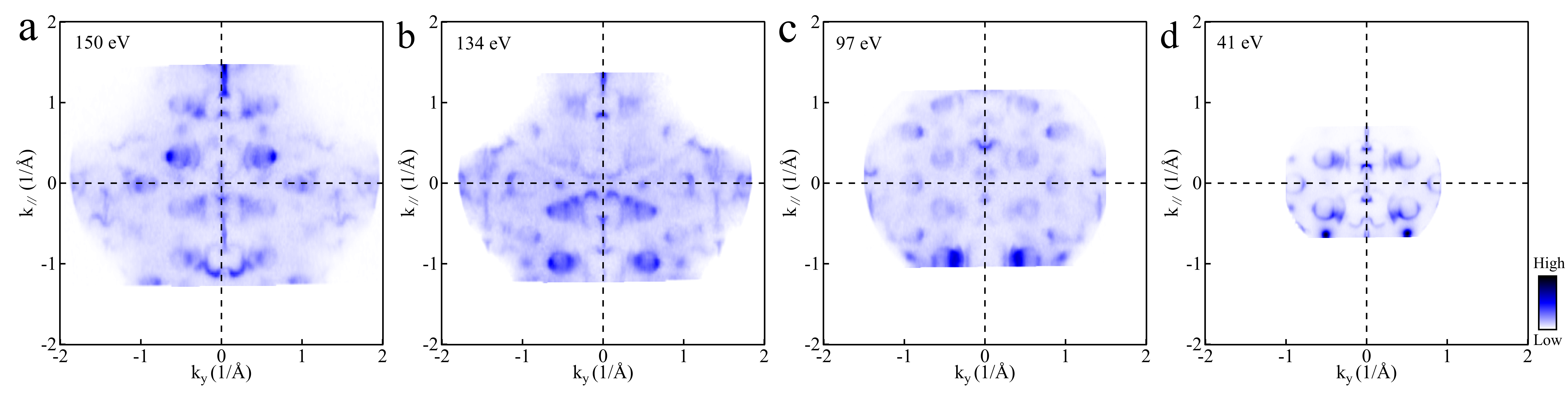}
\end{center}
\caption{\footnotesize\textbf{Photon energy dependent Fermi surface of NdAlSi (103) surface.} (a-d) Fermi surfaces of NdAlSi (103) surface measured on a flat (well defined) region at different photon energies: (a) 150 eV, (b) 134 eV, (c) 97 eV, and (d) 41 eV.
}
\label{S1}
\end{figure*}

To verify the origin of the observed states, we systematically measured the Fermi surface of NdAlSi (103) surface using different photon energies, as shown in Fig.~\ref{S1}. The persistence of identical features across photon energies confirms that the observed signals originate from surface states (SSs) rather than bulk states (BSs). This establishes that the Fermi surface shown in Fig. 1e in the main text are indeed intrinsic to the SS of (103) surface.

\begin{figure*}[tbp]
\begin{center}
\includegraphics[width=1\columnwidth,angle=0]{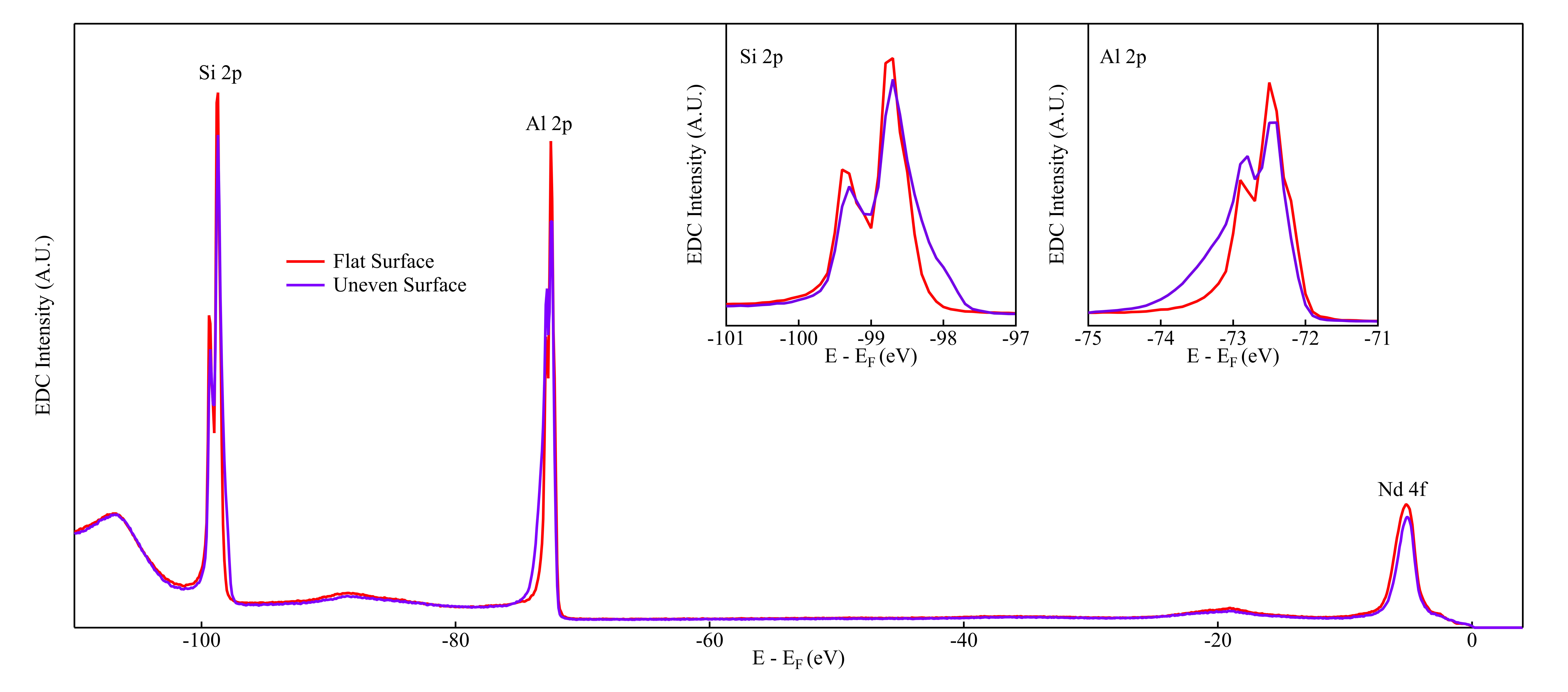}
\end{center}
\caption{\textbf{XPS of NdAlSi.} Comparison of core-level spectra from flat (red) and uneven (purple) surfaces. Insets show zoomed Si 2p and Al 2p peaks.
}
\label{S2}
\end{figure*}

Figure~\ref{S2} presents the X-ray photoemission spectroscopy (XPS) spectra measured with a photon energy of 200 eV, showing nearly identical core-level features for both flat and uneven surfaces, consistent with an unchanged chemical termination. A weak additional shoulder peak appears near the Si 2p core level on the uneven surface, indicating slight modifications of the local chemical environment that may arise from surface disorder. Nevertheless, the overall spectral shape and peak positions remain essentially the same. These results confirm that the suppression of SSs on uneven surfaces (Fig. 1f in the main text) is not due to a change in chemical termination, but rather originates from disorder-induced suppression of coherent SSs. Subsequent photon energy dependent measurements further confirm the bulk nature of the Fermi surface shown in Fig. 1e in the main text.\\

{\bf 2. Polarization dependent SS and dichroism measurements of NdAlSi (103) surface}

Figures~\ref{S3}a-\ref{S3}d display the measured Fermi surfaces of the NdAlSi (103) surface under linear horizontal (LH), linear vertical (LV), positive circular (PC), and negative circular (NC) polarizations. The overall features remain robust across different incident polarizations, reflecting their intrinsic electronic origin. Figs.~\ref{S3}e and~\ref{S3}f present the corresponding linear dichroic (LD = (LH–LV)/(LH+LV)) and circular dichroic (CD = (PC–NC)/(PC+NC)) Fermi surfaces, respectively. The strong dichroic contrast highlights orbital-selective contributions and provides an additional handle for disentangling SSs from bulk bands, as well as for tracking the evolution of topological SSs.\\

{\bf 3. Bulk Fermi surfaces of NdAlSi (103) surface}

Figures~\ref{S4}–\ref{S6} compare photon energy dependent bulk Fermi surface measurements (Fig.\ref{S4}) with DFT calculations of momentum-space slices along [103] (Fig.\ref{S5}) and perpendicular to [103] (Fig.~\ref{S6}). The measured Fermi surfaces exhibit pronounced out-of-plane dispersion, with spectral features shifting systematically as photon energy varies clear evidence of bulk-derived states. The corresponding DFT slices taken along and perpendicular to the [103] direction capture the same overall topology of the Fermi-surface pockets, their relative distribution in the surface plane, and the periodic shifts with changing k$_{\perp}$ and k$_{//}$. The close correspondence between experiment and theory confirms the bulk origin of these features and further illustrates the quasi-periodic in-plane translations of bulk projections that are essential for understanding the superlattice periodicity discussed in the main text.\\

{\bf 4. Commensurability of bulk projections with surface Brillouin zone periodicity}

The periodicity of the two-dimensional surface Brillouin zone (SBZ) is determined by the primitive in-plane reciprocal vectors associated with a given facet. To analyze the relation between bulk and surface periodicities, let $\{\mathbf b_1,\mathbf b_2,\mathbf b_3\}$ denote the primitive reciprocal vectors of the bulk lattice with metric tensor
\[
M_{ij}=\mathbf b_i\!\cdot\!\mathbf b_j,
\]
which encodes the lattice constants and inter-axial angles. For a surface defined by Miller indices $(hkl)$, the bulk reciprocal vector normal to the plane is
\[
\mathbf G=h\,\mathbf b_1+k\,\mathbf b_2+l\,\mathbf b_3.
\]
The projector onto the surface plane is
\[
P=\mathbb I-\frac{\mathbf G\otimes\mathbf G}{\mathbf G\!\cdot\!\mathbf G}.
\]
We choose the surface-BZ primitive basis as
\[
\mathbf g_1=P\mathbf b_1,\qquad \mathbf g_2=P\mathbf b_2 .
\]

Any bulk translation projected onto the surface, for example $P\mathbf b_3$, can be decomposed as
\begin{equation}
P\mathbf b_3=a_1\,\mathbf g_1+a_2\,\mathbf g_2 .
\label{eq:decompS6}
\end{equation}

\paragraph{From the decomposition to a linear system.}
Taking the inner product of Eq.~\eqref{eq:decompS6} with $\mathbf g_1$ and $\mathbf g_2$ gives
\begin{align}
\mathbf g_1\!\cdot\!(P\mathbf b_3) &= a_1(\mathbf g_1\!\cdot\!\mathbf g_1)+a_2(\mathbf g_1\!\cdot\!\mathbf g_2), \label{eq:S6p1}\\
\mathbf g_2\!\cdot\!(P\mathbf b_3) &= a_1(\mathbf g_2\!\cdot\!\mathbf g_1)+a_2(\mathbf g_2\!\cdot\!\mathbf g_2). \label{eq:S6p2}
\end{align}
Equations \eqref{eq:S6p1}–\eqref{eq:S6p2} are exactly the $2\times2$ linear system written in matrix form as

\begin{equation}
\underbrace{\begin{pmatrix}
\mathbf g_1\!\cdot\!\mathbf g_1 & \mathbf g_1\!\cdot\!\mathbf g_2\\
\mathbf g_2\!\cdot\!\mathbf g_1 & \mathbf g_2\!\cdot\!\mathbf g_2
\end{pmatrix}}_{\displaystyle G\ \text{(Gram matrix)}}
\begin{pmatrix} a_1\\[2pt] a_2 \end{pmatrix}
=
\begin{pmatrix}
\mathbf g_1\!\cdot\!(P\mathbf b_3)\\[2pt]
\mathbf g_2\!\cdot\!(P\mathbf b_3)
\end{pmatrix}.
\label{eq:2x2system}
\end{equation}

Because $\mathbf g_1,\mathbf g_2$ are linearly independent for any well-defined facet, the Gram matrix $G$ is symmetric positive-definite and hence invertible. Then, the coefficients $(a_1,a_2)$ are obtained by solving the $2\times 2$ linear system. 

\paragraph{Closed-form solution for $(a_1,a_2)$.}
Let
\[
\Delta=\det G=(\mathbf g_1\!\cdot\!\mathbf g_1)(\mathbf g_2\!\cdot\!\mathbf g_2)-(\mathbf g_1\!\cdot\!\mathbf g_2)^2>0 .
\]
Using Cramer's rule (or $G^{-1}$) yields
\begin{align}
a_1 &= \frac{(\mathbf g_2\!\cdot\!\mathbf g_2)\,\big[\mathbf g_1\!\cdot\!(P\mathbf b_3)\big]
      -(\mathbf g_1\!\cdot\!\mathbf g_2)\,\big[\mathbf g_2\!\cdot\!(P\mathbf b_3)\big]}{\Delta}, \label{eq:a1closed}\\[4pt]
a_2 &= \frac{(\mathbf g_1\!\cdot\!\mathbf g_1)\,\big[\mathbf g_2\!\cdot\!(P\mathbf b_3)\big]
      -(\mathbf g_1\!\cdot\!\mathbf g_2)\,\big[\mathbf g_1\!\cdot\!(P\mathbf b_3)\big]}{\Delta}. \label{eq:a2closed}
\end{align}
If an orthogonalized surface basis is chosen such that $\mathbf g_1\!\cdot\!\mathbf g_2=0$, Eqs.~\eqref{eq:a1closed}–\eqref{eq:a2closed} simplify to
\[
a_1=\frac{\mathbf g_1\!\cdot\!(P\mathbf b_3)}{\|\mathbf g_1\|^2},\qquad
a_2=\frac{\mathbf g_2\!\cdot\!(P\mathbf b_3)}{\|\mathbf g_2\|^2}.
\]

In the special case of an orthogonal bulk lattice (cubic, tetragonal, or orthorhombic), these coefficients can be written in closed form directly in terms of the Miller indices $(hkl)$. One finds
\[
a_1 = -\frac{h}{l}, \qquad a_2 = -\frac{k}{l}, \quad (l\neq 0),
\]
which shows that the in-plane displacement per bulk BZ step is controlled entirely by the ratios of Miller indices. If $l$ = 0, then we need to consider selecting other primitive basis vectors.

\paragraph{Commensurability along the out-of-plane stacking direction.}

 Here we ask whether the projected shift between successive bulk BZs along the surface normal closes after a finite number of repetitions. This amounts to finding the smallest $n\in\mathbb{N}$ such that
    \[
    n a_1 \in \mathbb{Z}, \quad n a_2 \in \mathbb{Z}.
    \]
    If $a_i=p_i/q_i\in\mathbb{Q}$ in lowest terms, define $\mathrm{den}(a_i)=q_i$; if $a_i=0$, then $\mathrm{den}(0)=1$. If either $a_i$ is irrational, $N_{\min}=\infty$. Otherwise,
    \[
    N_{\min}=\mathrm{lcm}\!\big[\mathrm{den}(a_1),\mathrm{den}(a_2)\big].
    \]
    Here, lcm denotes the least common multiple (LCM). With $\mathrm{den}(0)=1$, the definition remains consistent for zero, integer, and rational coefficients, while irrational values yield $N_{\min}=\infty$. Thus, $N_{\min}$ reflects only the closure of the one-dimensional stacking sequence.

\paragraph{Minimal surface repeat unit for computation.}
An immediate corollary of the above is that the surface cell enlarged by $N_{\min}$ along directions where $a_i$ has a nontrivial denominator is exactly the \emph{minimal surface repeat unit} commensurate with bulk projections. This cell is therefore the correct choice for DFT slab or surface Green’s-function calculations. Any smaller cell would fail to satisfy the bulk–surface commensurability condition and thus misrepresent the surface-state periodicity.

More explicitly, the logic is as follows: each successive bulk BZ projects into the SBZ displaced by $a_1\mathbf g_1+a_2\mathbf g_2$. If $a_i$ are rational, then after $N_{\min}$ repetitions the total displacement equals an integer linear combination of $\mathbf g_1$ and $\mathbf g_2$. This integer relation means that all distinct projections are now identified modulo the SBZ lattice. Hence, extending the surface unit cell by $N_{\min}$ along those directions generates a surface reciprocal lattice that exactly matches the closure of the projected bulk sequence. By contrast, choosing a smaller cell ($n<N_{\min}$) would imply that $n a_1,n a_2\in \mathbb Z$, which contradicts the minimality of $N_{\min}$. Therefore, the $N_{\min}$-extended unit cell is not just sufficient but also necessary.

In practice, this prescription is straightforward: 
\begin{enumerate}
    \item Examine the coefficients $(a_1,a_2)$;
    \item If either is fractional, enlarge the corresponding real-space lattice vector by the least common multiple $N_{\min}$;
    \item Construct the slab using this enlarged cell.
\end{enumerate}
The resulting SBZ is guaranteed to be the smallest one consistent with bulk periodicity. This ensures that calculated surface spectral functions will display the correct band folding and SSs as observed in ARPES, without artificial replicas.

\paragraph{Finite-cutoff matching between adjacent SBZs.}

The geometric commensurability discussed above (with period $N_{\min}=\mathrm{lcm}[\mathrm{den}(a_1),\mathrm{den}(a_2)]$) guarantees that the \emph{translation class} of bulk-to-surface projections closes after $N_{\min}$ bulk BZs, i.e.\ $N_{\min}\,P\mathbf b_\perp\in\Lambda_{\rm SBZ}$. 
However, when one compares the \emph{finite-sum} projection obtained by summing only over a truncated set of bulk zones $\Omega=\{1,2,\dots,M\}$, the content inside two adjacent SBZs generally differs unless the index set $\Omega$ is invariant under the shift $m\mapsto m+N_{\min}$ induced by moving from SBZ$_n$ to SBZ$_{n+1}$. 
Formally, if $S(\Omega)$ denotes the projected set in SBZ$_1$ from indices $\Omega$, then the projected set in SBZ$_2$ is $S(\Omega+N_{\min})$ with $\Omega+N_{\min}=\{m+N_{\min}\,|\,m\in\Omega\}$. 
To have \emph{exact} equality $S(\Omega)=S(\Omega+N_{\min})$ at finite cutoff, one needs $\Omega=\Omega+N_{\min}$.  
The smallest nonempty finite set with this property is two consecutive periods:
\[
\Omega_{\min}=\{1,2,\dots,2N_{\min}\},
\]
so the minimal cutoff ensuring equality between SBZ$_1$ and SBZ$_2$ under a finite projection is
\[
M_{\min}^{(\mathrm{adj})}=2N_{\min}.
\]
In the infinite-sum limit ($\Omega=\mathbb Z$) this equality holds trivially.\\

\textbf{Two notions of commensurability.}  
It is important to distinguish two related but not identical concepts:

\begin{enumerate}
    \item \emph{Geometric commensurability.}  
    The parameter $N_{\min}$ reflects the closure of the one-dimensional stacking sequence defined by the projected bulk reciprocal lattice vectors. This purely geometric treatment considers only the relation between $P\mathbf b_i$ and the primitive basis of the SBZ, and establishes whether the projected periodicities close after a finite number of bulk BZs (commensurate case) or only in the infinite limit (incommensurate case). 
    
    \item \emph{Spectral commensurability.}     
    In practice, the bulk electronic structure--such as the distribution of Weyl points or band extrema--does not necessarily preserve the full symmetry of the BZ. Consequently, projections from higher-order bulk zones are not strictly equivalent to those from the first, and additional zones (as discussed above, typically up to $2N_{\min}$) are required to reproduce even partial spectral equivalence. Moreover, because the bulk reciprocal lattice generates the entire set $\{m_1\mathbf b_1+m_2\mathbf b_2+m_3\mathbf b_3\}$, all of which project onto the surface, exact reconstruction of the full SBZ periodicity requires contributions from \emph{infinitely many} bulk BZs. Thus, no finite cutoff can fully capture all projected Weyl points and surface-state replicas across the two-dimensional SBZ.
\end{enumerate}

\subsection*{Example: (103) surface of tetragonal NdAlSi}

For $(h,k,l)=(1,0,3)$ in tetragonal NdAlSi ($a=b=4.204$ \AA, $c=14.519$ \AA\cite{CLi_AM2025_OTjernberg}) one obtains
\[
P\mathbf b_3 = -\tfrac{1}{3}\,\mathbf g_1 + 0\cdot \mathbf g_2,
\]
so $a_1=-1/3$, $a_2=0$, and hence 
\[
N_{\min}=\mathrm{lcm}(3,1)=3,
\]
independent of the specific values of $a$ and $c$. Thus, along the out-of-plane stacking direction, the projected shift closes exactly after three bulk BZs ($N_{\min}=3$). This means that each successive bulk BZ projects onto the (103) plane as a lateral shift of $-1/3 g_1$. As a result, the first, second, and third bulk BZs contribute distinct projections within the first SBZ, while the fourth bulk BZ closes the cycle by returning to a displacement of $-g_1$, which is equivalent to the first.\\

\emph{Worked application of the prescription.}  
Applying the minimal repeat unit rule:
\begin{itemize}
    \item \textbf{Step 1 (identify fractional directions):} $a_1=-1/3$ is fractional, while $a_2=0$ is integer.
    \item \textbf{Step 2 (build the minimal cell):} enlarge the real-space surface lattice vector dual to $\mathbf g_1$ by a factor of 3; keep the vector dual to $\mathbf g_2$ primitive. The slab in-plane cell is therefore $3\times1$ with respect to the primitive (103) surface lattice.
    \item \textbf{Step 3 (compute):} construct a slab using this $3\times1$ in-plane cell, with sufficient thickness and vacuum. For Wannier-based semi-infinite Green’s functions, choose a principal layer compatible with the $3\times1$ periodicity. The resulting surface spectral function $A(\mathbf k_{\parallel},\omega)$ reproduces the experimentally observed SBZ closure and the correct placement of SSs; a primitive ($1\times1$) cell would instead yield mismatches and spurious folding.
\end{itemize}

\noindent\textbf{Notes.}
 
(i) If multiple chemical terminations exist, compute each with the same $3\times1$ periodicity and compare or mix them according to experiment.  

(ii) Weak surface relaxation does not alter $N_{\min}$ (a property of the metric/projection), but can slightly shift surface bands; relaxing only the top few layers is usually sufficient.  

(iii) If a true real-space reconstruction exists (e.g., $5\times1$), that reconstruction--not the LCM cycle--sets the minimal computational cell.\\

\emph{Remark on conventional fcc indexing.} Since NdAlSi is \emph{body-centered cubic (bcc)} in real space, its reciprocal lattice is \emph{face-centered cubic (fcc)}. Under the conventional fcc indexing rule (all-even/all-odd $hkl$), the nearest allowed reciprocal point along $[001]$ is $(002)$ rather than $(001)$. Consequently, the bulk first-BZ half-width along the surface normal ($k_z$) doubles (e.g.\ from $\pm 0.2164$ to $\pm 0.4328$~\AA$^{-1}$, total $0.8656$~\AA$^{-1}$). This change affects only the normal direction and does not alter the in-plane SBZ' size ($\mathbf g'_1,\mathbf g_2$). In this convention, adjacent bulk BZs separated by $2\mathbf b_3$ yield an in-plane shift
\[
P(2\mathbf b'_3)=2\,P\mathbf b'_3=-\tfrac{2}{3}\,\mathbf g'_1,
\]
which is equivalent to the primitive-basis statement $P\mathbf b'_3=-\tfrac{1}{3}\,\mathbf g'_1$ and does not modify $N_{\min}$. Equivalently, the minimal commensuration along the $\mathbf g'_1$ direction reads
\[
\,3\times\Big(\tfrac{2}{3}\,\mathbf g'_1\Big)=2\,\mathbf g'_1\,
\]
i.e. three bulk projections commensurate with two $2\,\mathrm{SBZ}'$. Hence, in this conventional fcc enumeration the minimal in-plane supercell spans $2\,\mathrm{SBZ}'$ widths along $\mathbf g'_1$ (and $1$ along $\mathbf g_2$). To keep the apparent three-to-one commensuration explicit under the conventional fcc notation, we \emph{identify} $2\,\mathrm{SBZ}'$ as the (redefined) $\mathrm{SBZ}$ and set $\mathbf g_{1}\equiv 2\mathbf g'_1$. In this convention, the bulk primitive basis vectors become $\mathbf b_{1}\equiv 2\mathbf b'_1$ and $\mathbf b_{3}\equiv 2\mathbf b'_3$. With this unified notation, the statement “three bulk projections match one surface unit’’ remains valid.

\noindent\textbf{Notes.} According to the same convention (i.e., the conventional fcc indexing rule requiring all-even or all-odd $hkl$), the nearest allowed reciprocal point along $[001]$ is $(002)$ rather than $(001)$. Strictly speaking, and for the sake of notational consistency, this implies that the SBZ should also be doubled along $\mathbf g'_2$ (i.e., $\mathbf g_2 \equiv 2\mathbf g'_2$). However, since this direction is commensurate, we do not elaborate further so as to keep the focus on the incommensurate direction, whose treatment already captures the essential physics and avoids additional notational complexity.\\

\emph{Layer-changing second-BZ neighbors.}
Defining the “second-order BZ’’ as all six neighbors adjacent to the first (labels \#1–\#6 in Fig.~\ref{S7}a), their surface projections populate the in-plane displacements
\[
\{\ \pm\tfrac{1}{3}\,\mathbf g_1,\ \pm\tfrac{2}{3}\,\mathbf g_1\ \}\ \oplus\ \mathbb{Z}\,\mathbf g_2.
\]
Thus \#1 and \#5 map to $+\tfrac{1}{3}\,\mathbf g_1$, \#2 and \#6 to $-\tfrac{1}{3}\,\mathbf g_1$, while \#3 and \#4 correspond to $\pm\tfrac{2}{3}\,\mathbf g_1$. The latter coincide in position with third–BZ copies at $\pm\tfrac{2}{3}\,\mathbf g_1$, explaining the diagonal offsets visible in Fig.~\ref{S7}a. 

It is important to note, however, that such “position coincidence’’ does not imply full spectral equivalence. The second-order neighbors do not reproduce the multiplicity and symmetry of contributions from higher orders: e.g., the $\pm\tfrac{2}{3}\,\mathbf g_1$ shifts appear only once in the second-order set, whereas the corresponding $\pm\tfrac{2}{3}\,\mathbf g_1$ displacements in the third-order set occur in a different multiplicity pattern. Consequently, the projected features inside SBZ$_1$ and SBZ$_2$ are still not identical if only the second-order BZs are included. In other words, although some higher-order points are “mimicked’’ in position by lower-order neighbors, their relative weights and degeneracies remain unmatched.

Geometric closure remains $N_{\min}=3$, reflecting that three successive bulk projections along the (103) normal return to an equivalent in-plane shift. To enforce strict visual identity between adjacent SBZs under a finite projection, one must adopt a symmetric cutoff that includes both positive and negative orders up to
\[
M_{\min}^{(\mathrm{adj})}=2N_{\min}=6.
\]
The distinction between the geometric cycle ($N_{\min}=3$) and the finite-cutoff condition ($M_{\min}=6$) accounts for this observation. As one proceeds to even higher orders, the set of contributing bulk BZs expands further; no finite cutoff can reproduce the full spectral content exactly. Only in the infinite limit does the superlattice generated by all bulk BZ projections become exactly commensurate with the SBZ. This is why the surface periodicity is progressively restored as more bulk BZs are added, while exact closure is only reached in the infinite-projection limit.

\subsection*{Example: (001) surface of tetragonal NdAlSi}

For $(h,k,l)=(0,0,1)$ one finds $P\mathbf b_3=0$, so $a_1=a_2=0$ and $N_{\min}=1$. Both geometric and spectral commensurability are satisfied immediately, and already the first bulk BZ reproduces the full surface periodicity.

\subsection*{Remarks on non-orthogonal lattices}

For non-orthogonal systems (e.g.\ hexagonal, rhombohedral, monoclinic), the coefficients $(a_1, a_2)$ depend explicitly on ratios such as $a/c$ and interaxial angles. In such cases $a_1$ or $a_2$ may be irrational, leading to $N_{\min}=\infty$ even for the one-dimensional stacking sequence. This highlights the genuine metric dependence of the commensurability problem.

\subsection*{Visualization of bulk BZ projections on the (103) surface}

Fig.~\ref{S7} illustrates how bulk BZs project onto the low-symmetry (103) surface and generate an emergent superlattice periodicity. Fig.~\ref{S7}a shows the projection of Weyl points from the first and six second-order BZs onto the (103) plane. Due to the incommensurability between bulk reciprocal vectors and the (103) surface basis, consecutive BZ projections appear shifted within the surface plane, rather than coinciding as on high-symmetry surfaces. Fig.~\ref{S7}b presents a schematic stacking view: successive second-order BZs project as laterally displaced replicas, and their superposition yields a new periodicity--a surface-consistent superlattice (a momentum-space moir$\mathrm{\acute{e}}$ modulation). This visualization highlights how including higher-order bulk BZ projections restores the apparent mismatch between bulk and surface periodicities, thereby resolving the bulk-boundary correspondence (BBC) paradox on low-symmetry facets.

To further illustrate this mechanism, we explicitly integrate the bulk spectral weight from first-principles calculations over finite windows in momentum space. Fig.~\ref{S8}a shows the Fermi surface obtained by integrating planes parallel to the (103) surface within $k_{\perp}=0$-$0.5$ Å$^{-1}$, while Fig.~\ref{S8}b presents the result for perpendicular planes within $k_{//}=0$-$1.5$ Å$^{-1}$. For small integration ranges, the projected bulk features appear misaligned with the surface periodicity. As the integration window expands, contributions from multiple bulk BZs are superimposed, and the apparent periodicity progressively approaches that of the (103) SBZ. This provides a direct visualization of how higher-order bulk projections gradually restore the correct surface periodicity, consistent with the generalized framework established above.

This generalized framework demonstrates that the apparent breakdown of BBC on low-symmetry surfaces is not a failure of topology but a consequence of incomplete bulk projection. By superimposing successive bulk BZs, a superlattice periodicity emerges that is fully commensurate with the SBZ, thereby restoring consistency between bulk and SSs. An immediate implication of this incommensurability is that topological surface Fermi arcs (SFAs) originating from different bulk zones may no longer project in registry. Instead, they overlap and hybridize on the surface, leading to new boundary features that have no analogue on high-symmetry facets. In particular, such hybridization provides a natural pathway for the formation of closed surface Fermi arc loops (SFALs), which we discuss in detail later.\\

{\bf 5. Surface state calculations and comparisons}

Figure~\ref{S9} illustrates possible cleavage terminations that may arise when exposing the (103) surface of NdAlSi. Fig.~\ref{S9}a shows one possible cleavage mode that results in a zigzag-shaped (103) surface. This geometry can be understood as choosing an equivalent cleavage layer within each primitive unit cell. Since the primitive cells are translationally equivalent, cleavage is expected to occur along symmetry-equivalent layers, leading naturally to zigzag-type terminations. Fig.~\ref{S9}b exhibits an alternative description that emerges when three consecutive primitive cells are grouped into a supercell. Within this enlarged supercell, four nonequivalent cleavage layers exist, corresponding to eight distinct terminations of the (103) surface. The example in Fig.~\ref{S9}b illustrates one such construction, highlighting how the choice of supercell redefines the possible cleavage options. In this construction, the black parallelogram marks the minimal repeat unit of the (103) surface lattice. Because its spanning vectors are oblique, this cell is effectively monoclinic, despite the bulk crystal being body-centered tetragonal. This effective monoclinic character also shows why bulk and surface periodicities are intrinsically incommensurate when only a single surface cell is considered, and why the LCM criterion is required to restore commensuration. For the (103) surface of NdAlSi, this corresponds to a three-cell supercell that reconciles the SBZ with the projected bulk BZ.

In general, different choices of supercell lead to additional possibilities for the (103) cleavage geometry, and thus a wide variety of surface terminations may be realized. The actual cleavage plane adopted in experiment depends sensitively on the cleavage energy associated with each candidate layer: lower cleavage energy corresponds to a more favorable termination. Importantly, regardless of which cleavage mode is realized, a consistent description of the SSs requires that the LCM criterion be satisfied. Only by adopting a supercell commensurate with the LCM rule can the reconstructed SBZ and the associated topological SSs be described correctly.

Experimentally, ARPES measurements confirm the coexistence of multiple surface terminations. As shown in Fig.~\ref{S10}, besides the dominant Fermi surface structure in Fig.~\ref{S10}a, two additional distinct SS patterns were detected (Figs.~\ref{S10}b-\ref{S10}c). Together, these observations reveal at least three different SS configurations coexisting on the cleaved (103) surface. Importantly, these distinct spectra were obtained from different spatial regions of the same cleaved surface, demonstrating that a single macroscopic cleavage can naturally expose multiple microscopic terminations. Notably, despite their distinct dispersions, all three observed terminations exhibit the same reduced surface Brillouin zone periodicity, directly confirming that the LCM criterion governs SS commensuration independent of termination details. This experimentally confirms the universality of the LCM criterion established in Section 4. This consistency underscores the robustness of the LCM framework across different microscopic terminations. At the same time, we cannot exclude the possibility that even more variants may exist, consistent with the complexity expected for low-symmetry surfaces and the potential emergence of long-period zigzag-type reconstructions.

Based on this principle, the SS calculations shown in Fig.~\ref{S11} were carried out using the four possible cleavage layers identified within the three-cell supercell construction of Fig.~\ref{S9}b. This approach ensures that the computed surface spectra are both physically meaningful and directly comparable with experimental observations. Figs.~\ref{S11} and~\ref{S12} present the calculated surface Fermi surfaces for the (103) orientation of NdAlSi, considering four possible cleavage planes corresponding to eight distinct surface terminations. After extensive testing of these possibilities, we find that the calculation for a Si-terminated surface cleaved at the Al-Si layer (Fig.~\ref{S11}b) reproduces the experimentally measured dominant surface Fermi surface most accurately (Fig.~\ref{S10}a), confirming that the probed region corresponds to this termination.  Nonetheless, the additional surface structures observed in Figs.~\ref{S10}b-\ref{S10}c do not find convincing counterparts among the terminations considered here, indicating that their microscopic origin remains to be established. They are likely associated with more complex long-period zigzag-type cleavages that go beyond the four-layer options of the three-cell supercell in Fig.~\ref{S9}b, or with spatially mixed terminations within the probed regions.

Based on this principle, the SS calculations shown in Fig.~\ref{S11} were carried out using the four possible cleavage layers identified within the three-cell supercell construction of Fig.~\ref{S9}b. This approach ensures that the computed surface spectra are both physically meaningful and directly comparable with experimental observations. Figs.~\ref{S11} and~\ref{S12} present the calculated surface Fermi surfaces for the (103) orientation of NdAlSi, considering four possible cleavage planes corresponding to eight distinct surface terminations. After extensive testing of these possibilities, we find that the calculation for a Si-terminated surface cleaved at the Al–Si layer (Fig.~\ref{S11}b) provides the best agreement with the experimentally measured dominant surface Fermi surface (Fig.~\ref{S10}a), capturing its main features and indicating that the probed region most likely corresponds to this termination

Nonetheless, two important discrepancies remain. First, the additional surface structures observed in Figs.~\ref{S10}b–\ref{S10}c do not find convincing counterparts among the terminations considered here, indicating that their microscopic origin lies beyond the simple three-cell construction. They are likely associated with more complex long-period zigzag-type cleavages or with spatially mixed terminations within the probed regions. Second, even for the well-matched Si-terminated surface, the idealized calculations do not explicitly reproduce the closed connectivity of SFALs seen in ARPES. We attribute this mismatch to two factors: (i) the neglect of zigzag-type cleavages and mixed local terminations, which in real samples promote enhanced coupling between surface and bulk states; and (ii) the incomplete treatment of hybridization across multiple projected bulk BZs, whose lateral mismatch favors loop-like reconstructions of SSs. These considerations emphasize that while the LCM-consistent modeling captures the reduced periodicity and dominant spectral features, certain finer aspects of the connectivity remain emergent properties of real low-symmetry surfaces beyond the reach of ideal slab calculations.

Figure~\ref{S11} shows the results obtained using the Green's function method, which treats the system as semi-infinite and therefore automatically incorporates contributions from infinitely many bulk BZs. The output is a surface-projected spectral function. As a result, the “bulk” features appearing in these calculations are SBPSs, which inevitably couple to the boundary. When such projected BSs overlap with SSs in energy and momentum, they acquire mixed character--retaining bulk-like dispersion while carrying finite spectral weight at the surface. These mixed features correspond precisely to the SBRSs observed in ARPES.

By contrast, Fig.~\ref{S12} shows the results of a finite six-layer slab calculation in real space. Because the finite slab thickness quantizes the out-of-plane momentum $k_\perp$, only a discrete set of bulk states is sampled, which cannot reproduce the continuous bulk projections underlying genuine SBRSs. As a result, the bulk-like features are truncated and fail to hybridize correctly with the surface. Consequently, the agreement with experiment is poorer for the (103) surface, where the incommensurability between surface and bulk periodicities requires contributions from infinitely many bulk projections. Nevertheless, slab calculations remain useful for identifying SSs, and their degree of reliability depends strongly on the surface orientation.

This comparison highlights a fundamental distinction between low- and high-symmetry surfaces. For low-symmetry facets such as (103), Green's function calculations are indispensable for capturing both SSs and SBRSs, since only the semi-infinite approach properly restores the surface periodicity. In contrast, for high-symmetry surfaces such as (001), the bulk and SBZs are commensurate; in such cases, finite slabs already reproduce the correct periodicity, and while they do not describe resonance states, their calculated SSs often agree better with experiment than those from the Green's function method\cite{CLi_NC2023_OTjernberg}.

In this way, the present framework not only clarifies why Green’s function calculations succeed while finite slabs fail for low-symmetry surfaces, but also provides general guidance for future computational studies: slab models may suffice for high-symmetry facets, whereas low-symmetry surfaces intrinsically require semi-infinite Green’s function methods to faithfully capture their electronic structure.\\

{\bf 6. Identification and evolution of topological SFAs}

\textbf{Identification of SFAs} 

Figures~\ref{S13}a and~\ref{S13}b show the Fermi surfaces measured with LH and LV polarizations, with momentum cuts defined for further analysis. Figs.~\ref{S13}c-\ref{S13}f present dispersions along Cut1 and Cut2, which intersect the SFAs. The arc-like dispersions observed here are consistent with topological SFAs, as they do not close within a single SBZ. Figs.~\ref{S13}g-\ref{S13}j display dispersions along Loop1-Loop4 (L1-L4), which form closed contours in momentum space. In the conventional framework of BBC, the presence of a nonzero Chern number and projected chiral charge enclosed by such loops would provide unambiguous evidence for topological SFAs\cite{CLi_NC2023_OTjernberg,IBelopolski_PRL2016_MZHasan}.

To further clarify the SFA connectivity on the (103) surface, we performed semi-infinite Green's function calculations, as shown in Fig.~\ref{S14}. Fig.~\ref{S14}a presents the SBPSs, forming the continuum background. Fig.~\ref{S14}b shows the total surface spectrum, where both SBPSs and genuine SSs coexist. By subtracting the SBPS contribution in Fig.~\ref{S14}a, we obtain the pure SSs in panel Fig.~\ref{S14}c. Superimposed red and green markers indicate the projections of Weyl points from the first-order to third-order bulk BZs, with chirality $+1$ and $-1$, respectively. The comparison across Figs.~\ref{S14}a-\ref{S14}c highlights how projected Weyl points overlap with bulk resonances, how true SSs emerge on top of this continuum, and how the arc connectivity becomes evident once the pure surface contribution is isolated.  

Taken together, the experimental identification of arc-like dispersions in Fig.~\ref{S13}, combined with the theoretical separation of SBPSs and pure SSs in Fig.~\ref{S14}, provides a consistent picture: the observed arcs are topological SFAs connecting Weyl points of opposite chirality. While certain fine details differ between the ARPES spectra and the DFT-based surface calculations, the overall agreement is good. This correspondence supports the robustness of our assignment and provides a useful framework for interpreting the overall reconstruction of surface states. We note, however, that the detailed connectivity of SFAs in experiment is not fully captured by the idealized calculations, reflecting additional hybridization across multiple bulk projections and possible surface complexity beyond the simple termination model.\\

\textbf{Hybridization and emergence of SFALs} 

\begin{figure*}[tbp]
\begin{center}
\includegraphics[width=1\columnwidth,angle=0]{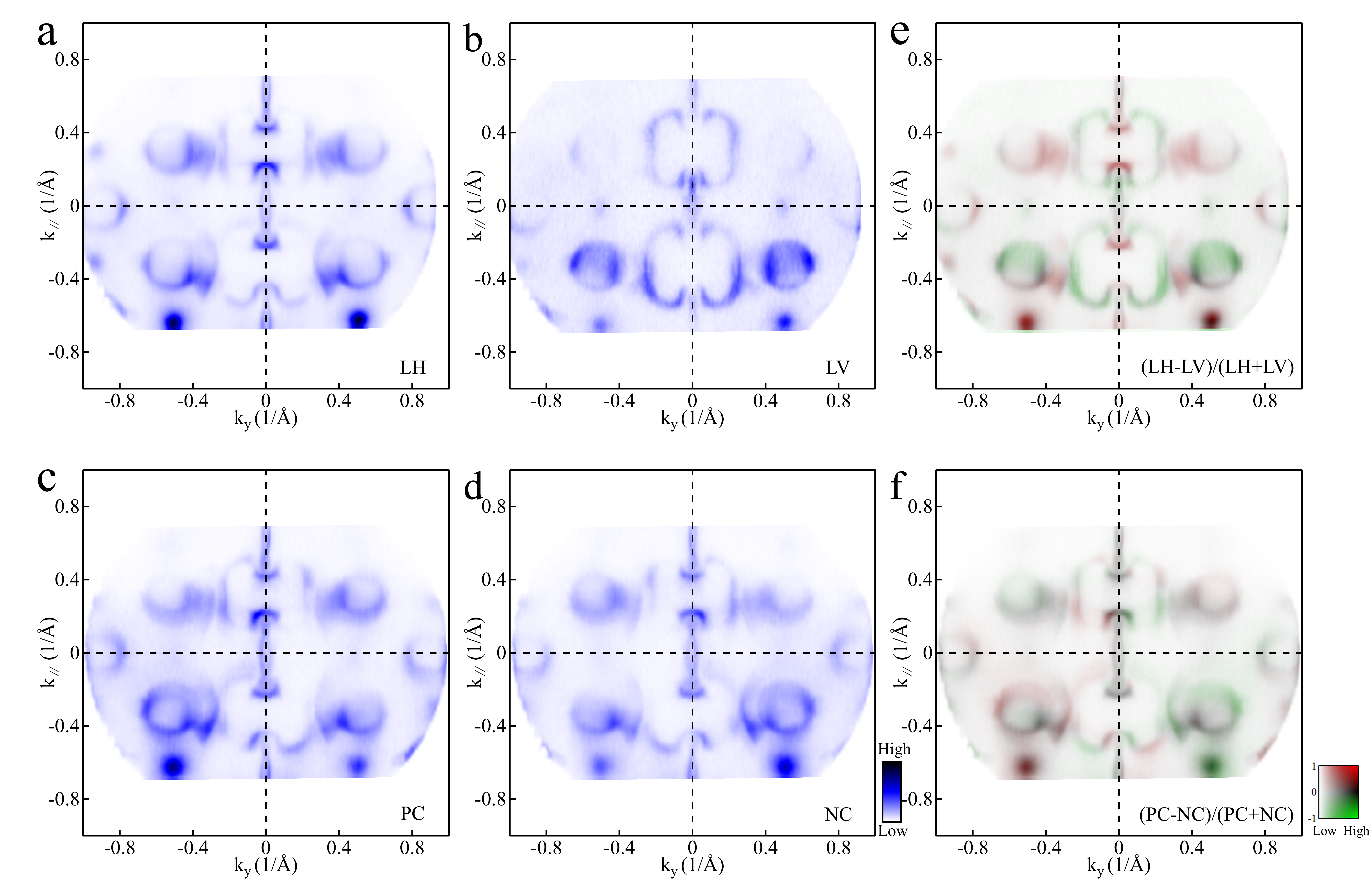}
\end{center}
\caption{\textbf{Polarization dependent SSs of NdAlSi (103) surface.} (a-d) Fermi surfaces of NdAlSi (103) surface measured on a well defined region at 41 eV photon energy under (a) linear horizontal (LH), (b) linear vertical (LV), (c) positive circular (PC), and (d) negative circular (NC) polarizations. (e) The linear dichroic (LD) Fermi surface obtained by the difference between (a) and (b). (f) The circular dichroic (CD) Fermi surface obtained by the difference between (c) and (d). The spectrum is plotted with the two-dimensional colorscale shown, with the horizontal axis corresponding to intensity and the vertical axis corresponding to LD or CD signal.
}
\label{S3}
\end{figure*}

\begin{figure*}[tbp]
\begin{center}
\includegraphics[width=1\columnwidth,angle=0]{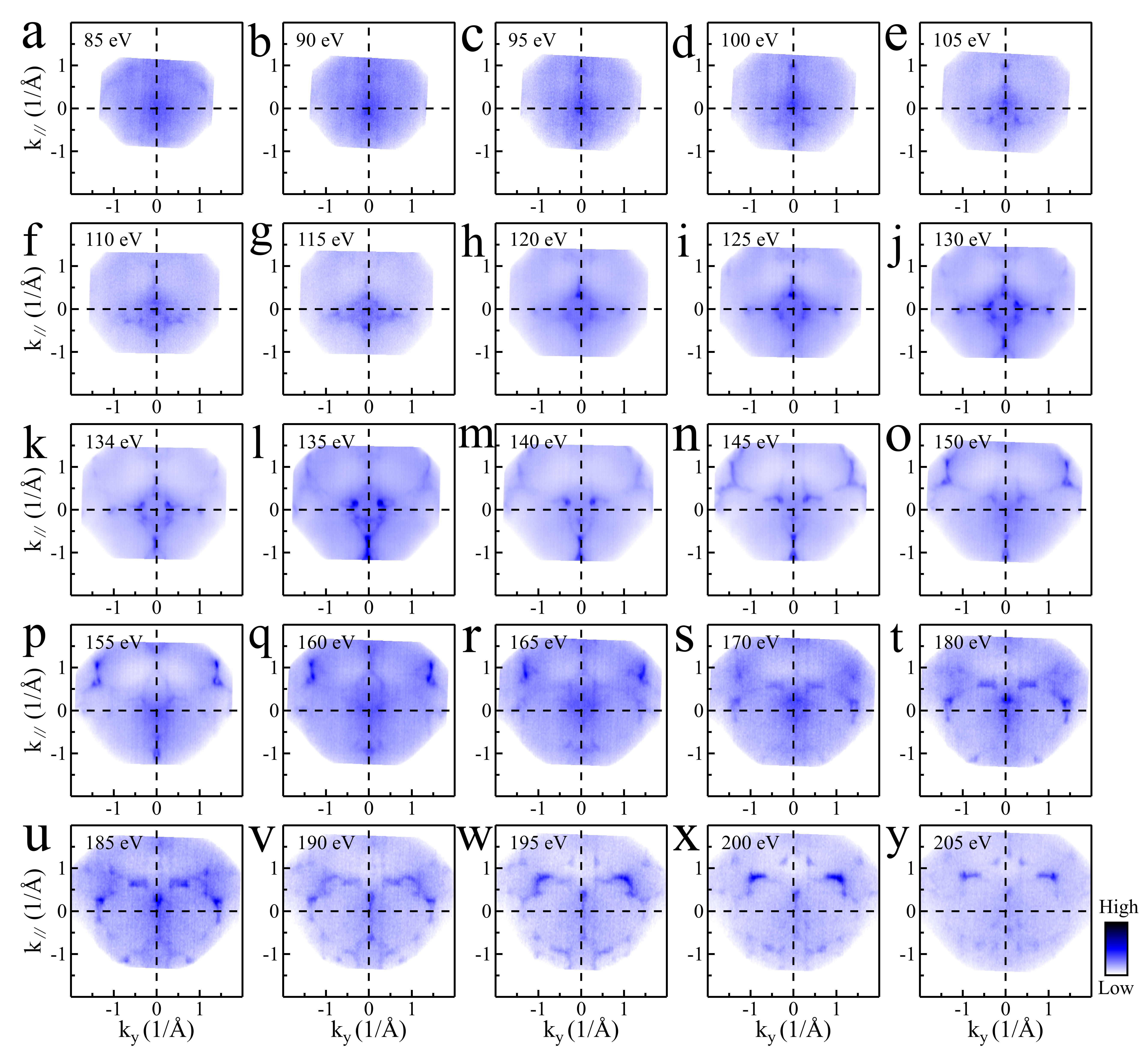}
\end{center}
\caption{\textbf{Photon energy dependent bulk projected Fermi surface on (103) surface.} (a-y) Bulk Fermi surfaces of NdAlSi (103) surface measured at photon energies from 85 eV to 205 eV.
}
\label{S4}
\end{figure*}

\begin{figure*}[tbp]
\begin{center}
\includegraphics[width=1\columnwidth,angle=0]{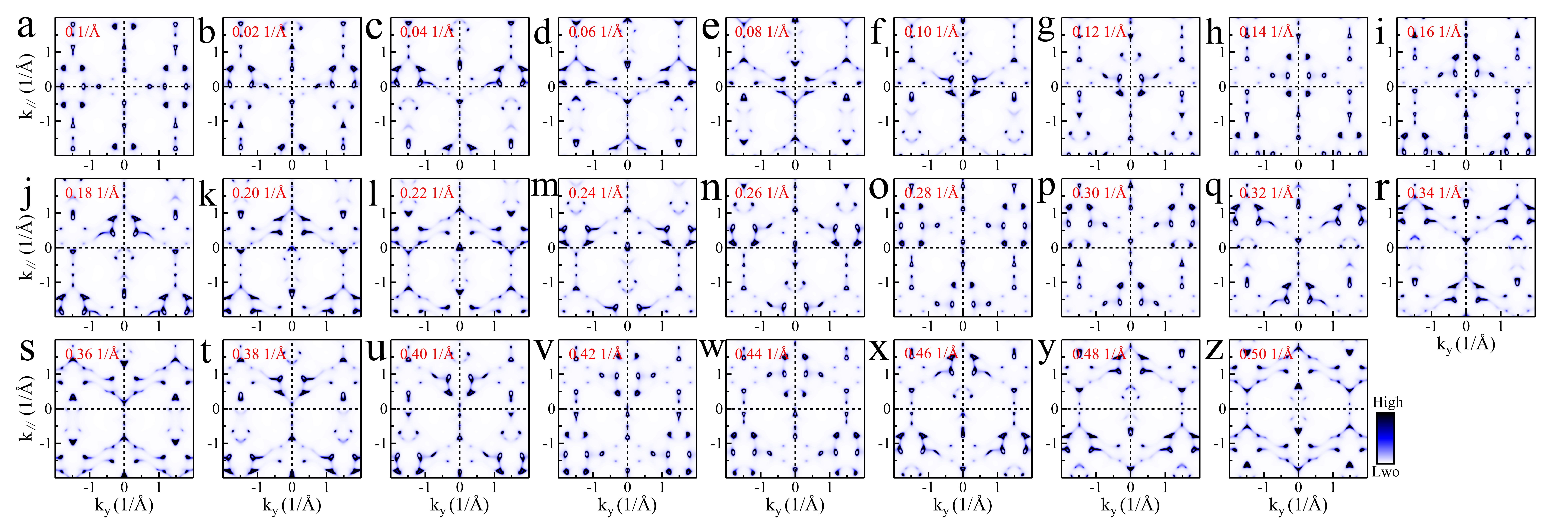}
\end{center}
\caption{(a-z) DFT-calculated bulk Fermi surface cross-sections along the [103] momentum direction.
}
\label{S5}
\end{figure*}

\begin{figure*}[tbp]
\begin{center}
\includegraphics[width=1\columnwidth,angle=0]{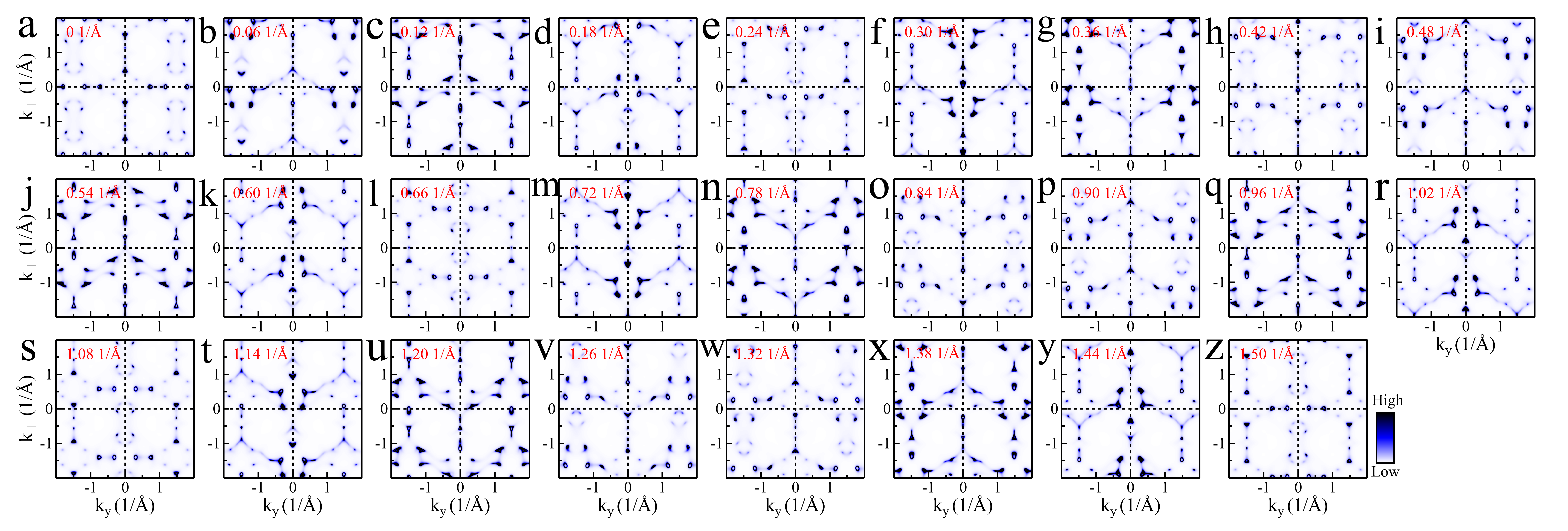}
\end{center}
\caption{(a-z) DFT-calculated bulk Fermi surface cross-sections taken along momentum directions perpendicular to [103].
}
\label{S6}
\end{figure*}

\begin{figure*}[tbp]
\begin{center}
\includegraphics[width=1\columnwidth,angle=0]{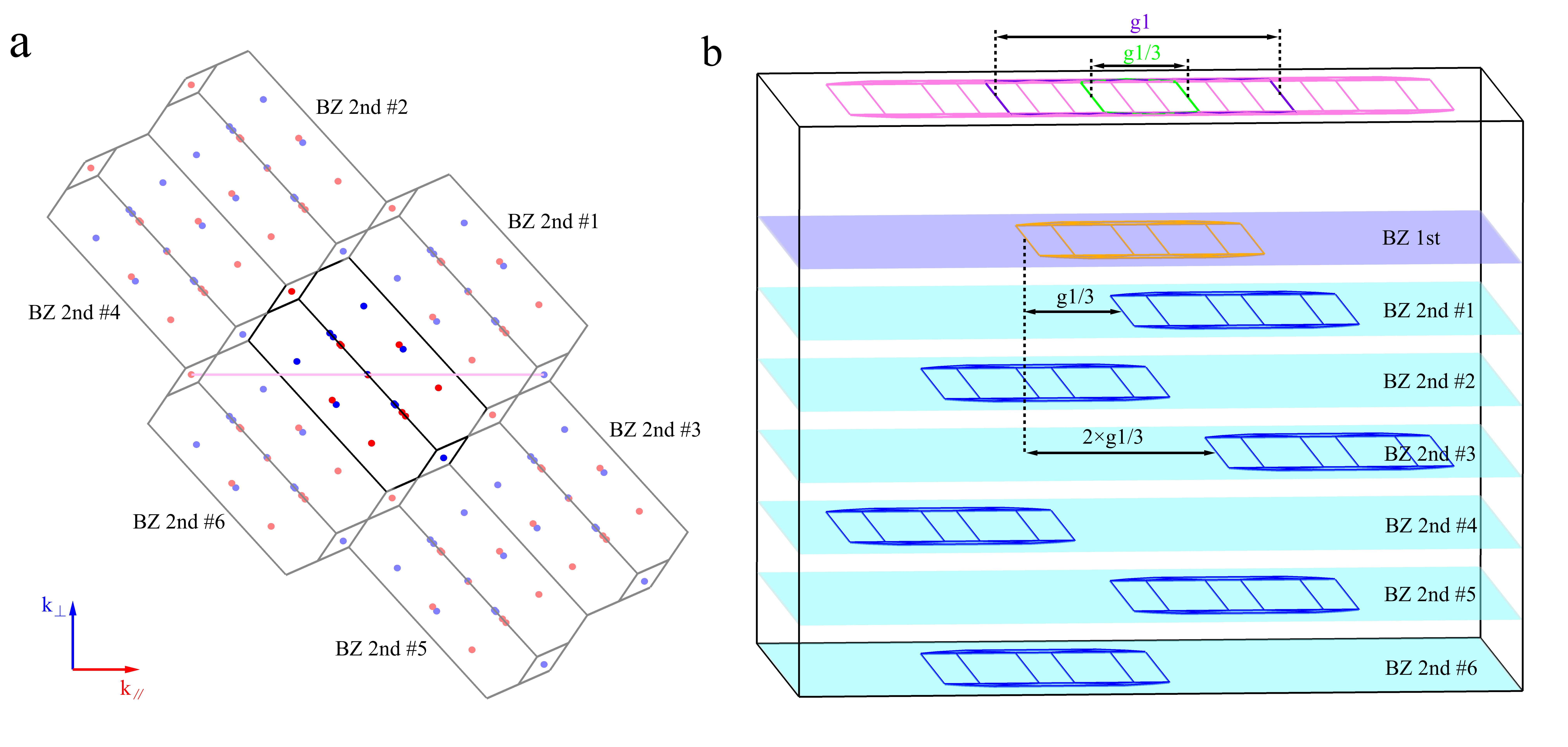}
\end{center}
\caption{\textbf{Projection of bulk BZ onto (103) surface.} (a) Projection of the first and six second-order bulk BZs onto the (103) plane. Successive second-order BZs project with in-plane displacements due to the incommensurability of bulk and surface reciprocal vectors. (b) Schematic stacking representation of the first (orange) and six second-order BZs (blue), projected onto the (103) surface. The displaced replicas accumulate into a superlattice periodicity that becomes commensurate with the SBZ, thereby restoring BBC on the (103) facet.
}
\label{S7}
\end{figure*}

\begin{figure*}[tbp]
\begin{center}
\includegraphics[width=1\columnwidth,angle=0]{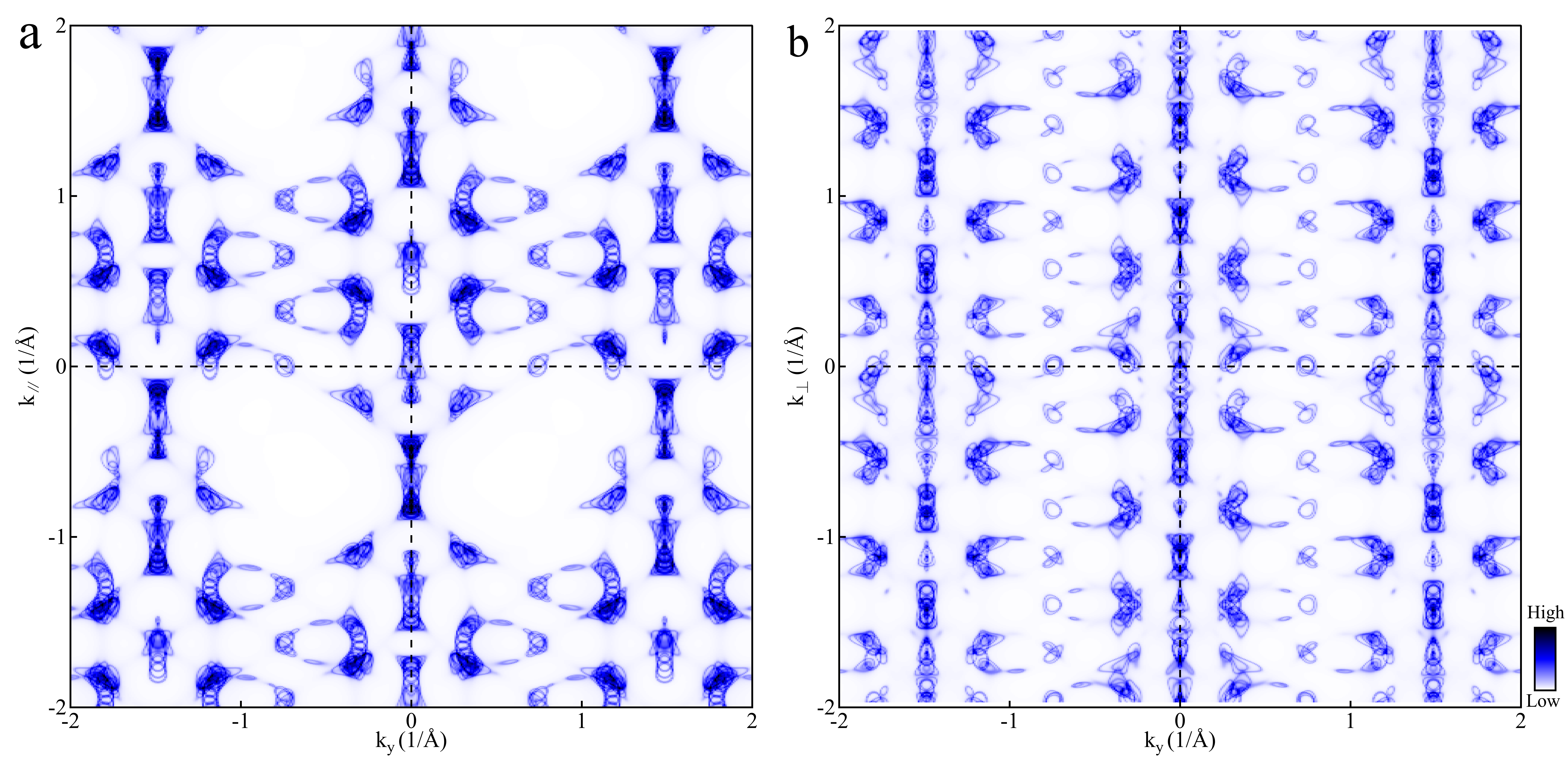}
\end{center}
\caption{\textbf{Bulk Fermi-surface projections and recovery of surface periodicity on the low-symmetry facet.} (a) Fermi surface obtained by integrating DFT-calculated bulk spectral weight over planes parallel to the (103) surface, with integration range $k_{\perp}=0$-$0.5$ Å$^{-1}$. This corresponds to a finite slice of momentum space along the direction perpendicular to the (103) facet. All the integrated data are taken from Fig.~\ref{S5}. (b) Same as (a), but integration is performed over planes perpendicular to the (103) surface, with range $k_{//}=0$-$1.5$ Å$^{-1}$, using data from Fig.~\ref{S6}. As the integration window increases, contributions from multiple bulk BZs are superimposed, and the apparent periodicity progressively converges toward that of the (103) SBZ.
}
\label{S8}
\end{figure*}

\begin{figure*}[tbp]
\begin{center}
\includegraphics[width=1\columnwidth,angle=0]{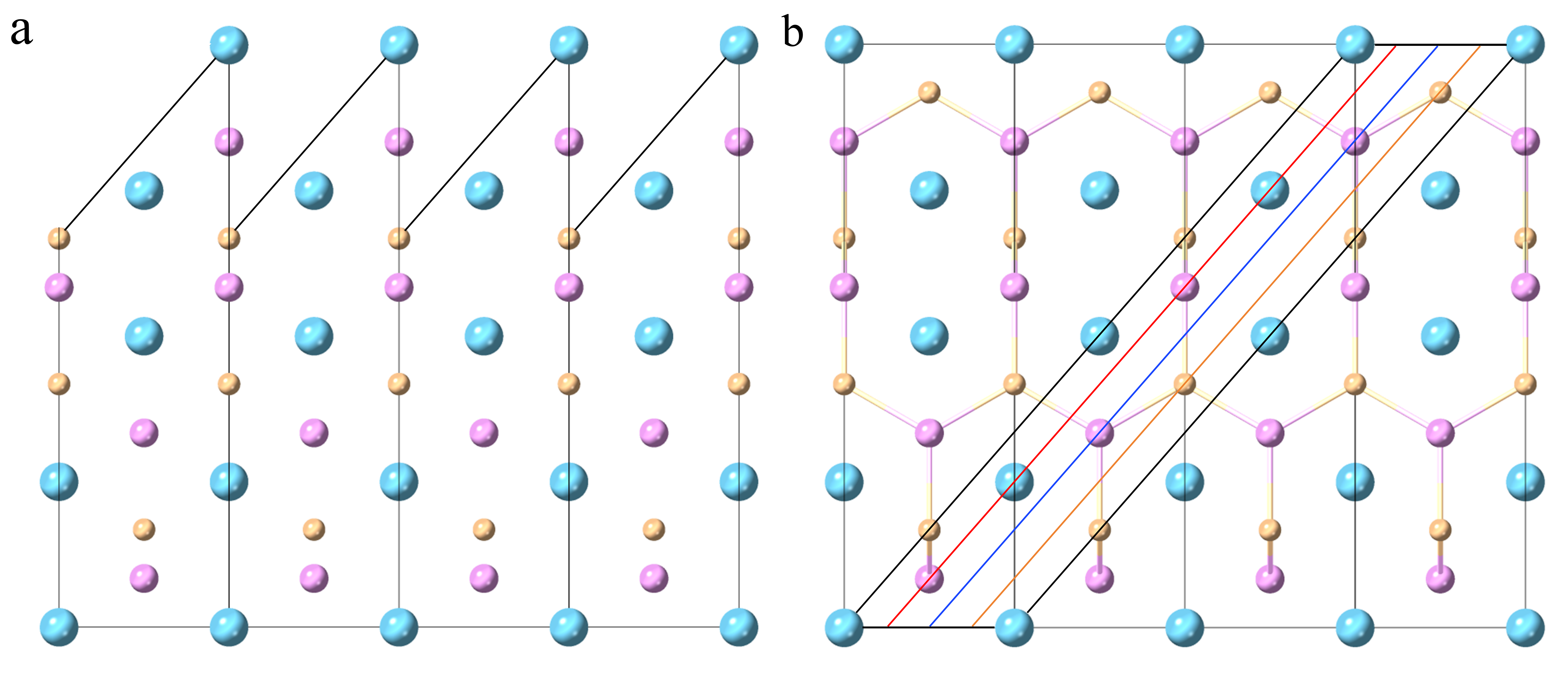}
\end{center}
\caption{\textbf{Possible cleavage geometries of the (103) surface.} (a) Example of a zigzag-shaped (103) cleavage obtained by cutting along equivalent layers in each primitive unit cell. (b) Cleavage constructed from a three-cell supercell, where four nonequivalent cleavage layers exist, yielding eight distinct terminations. While many cleavage geometries are possible in principle, a consistent description of the SSs requires satisfying the LCM criterion.
}
\label{S9}
\end{figure*}

\begin{figure*}[tbp]
\begin{center}
\includegraphics[width=1\columnwidth,angle=0]{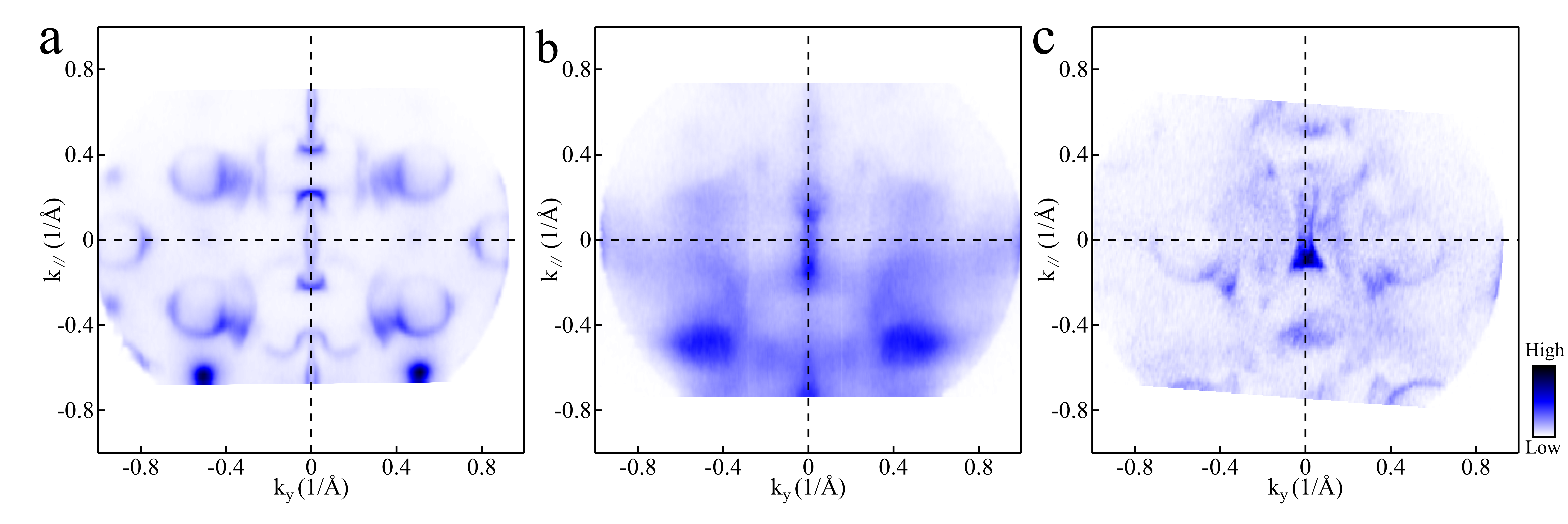}
\end{center}
\caption{\textbf{Multiple surface-state structures observed on the (103) surface of NdAlSi.} (a) Fermi surface structure corresponding to the termination discussed in the main text. (b-c) Two additional SS structures measured on different regions of the same (103) surface, corresponding to distinct cleavage terminations.
}
\label{S10}
\end{figure*}

\begin{figure*}[tbp]
\begin{center}
\includegraphics[width=1\columnwidth,angle=0]{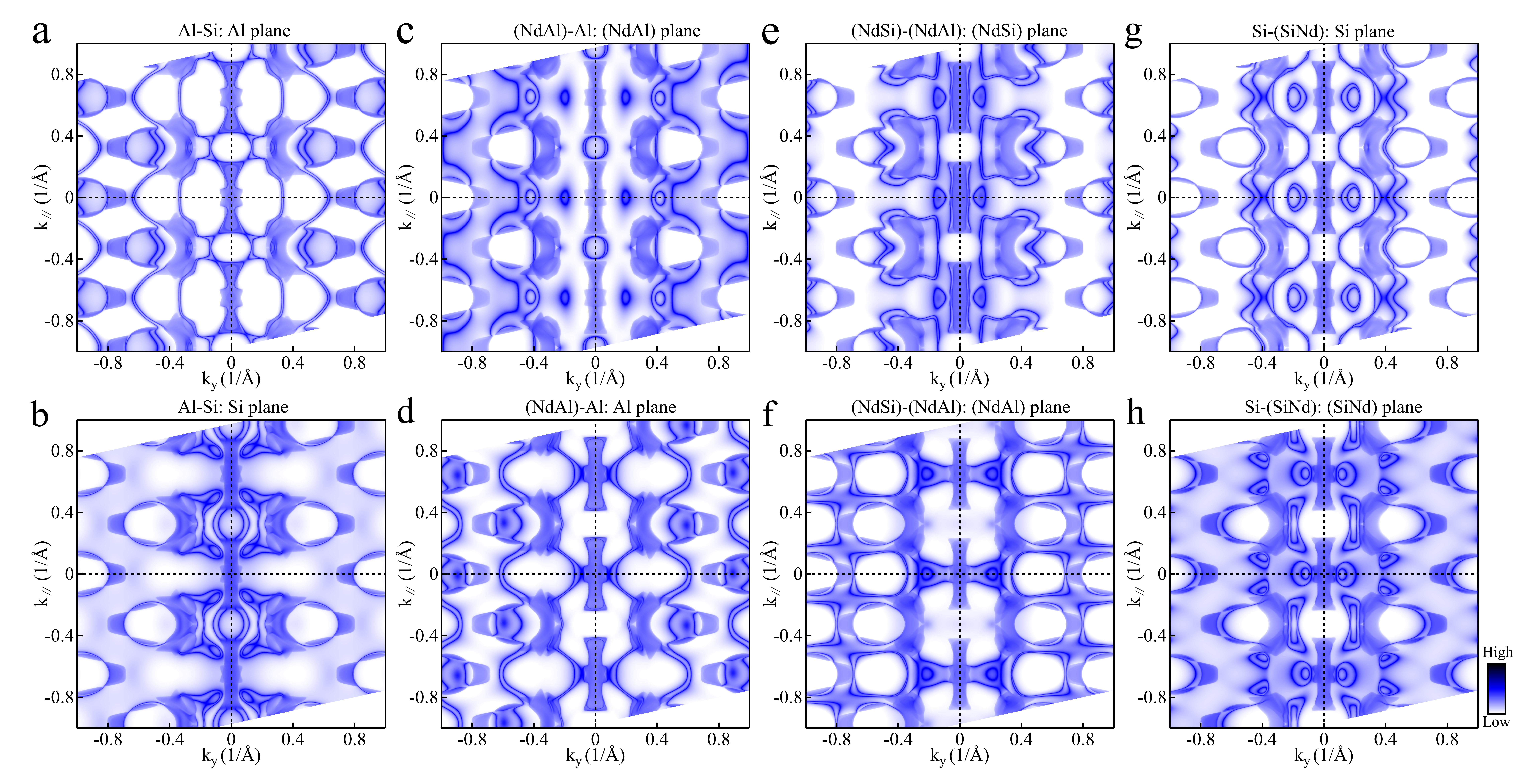}
\end{center}
\caption{Surface projected DFT calculations of (103) Fermi surfaces with semi-infinite Green's function for different cleavage terminations: (a) Al terminated surface cleaved at the Al-Si layer, (b) Si terminated surface cleaved at the Al-Si layer, (c) (NdAl) terminated surface cleaved at the (NdAl)-Al layer, (d) Al terminated surface cleaved at the (NdAl)-Al layer, (e) (NdSi) terminated surface cleaved at the (NdSi)-(NdAl) layer, (f) (NdAl) terminated surface cleaved at the (NdSi)-(NdAl) layer, (g) Si terminated surface cleaved at the Si-(SiNd) layer and (h) (SiNd) terminated surface cleaved at the Si-(SiNd) layer. The four cleavage layers defined within the supercell construction shown in Fig.~\ref{S9}b.
}
\label{S11}
\end{figure*}

\begin{figure*}[tbp]
\begin{center}
\includegraphics[width=1\columnwidth,angle=0]{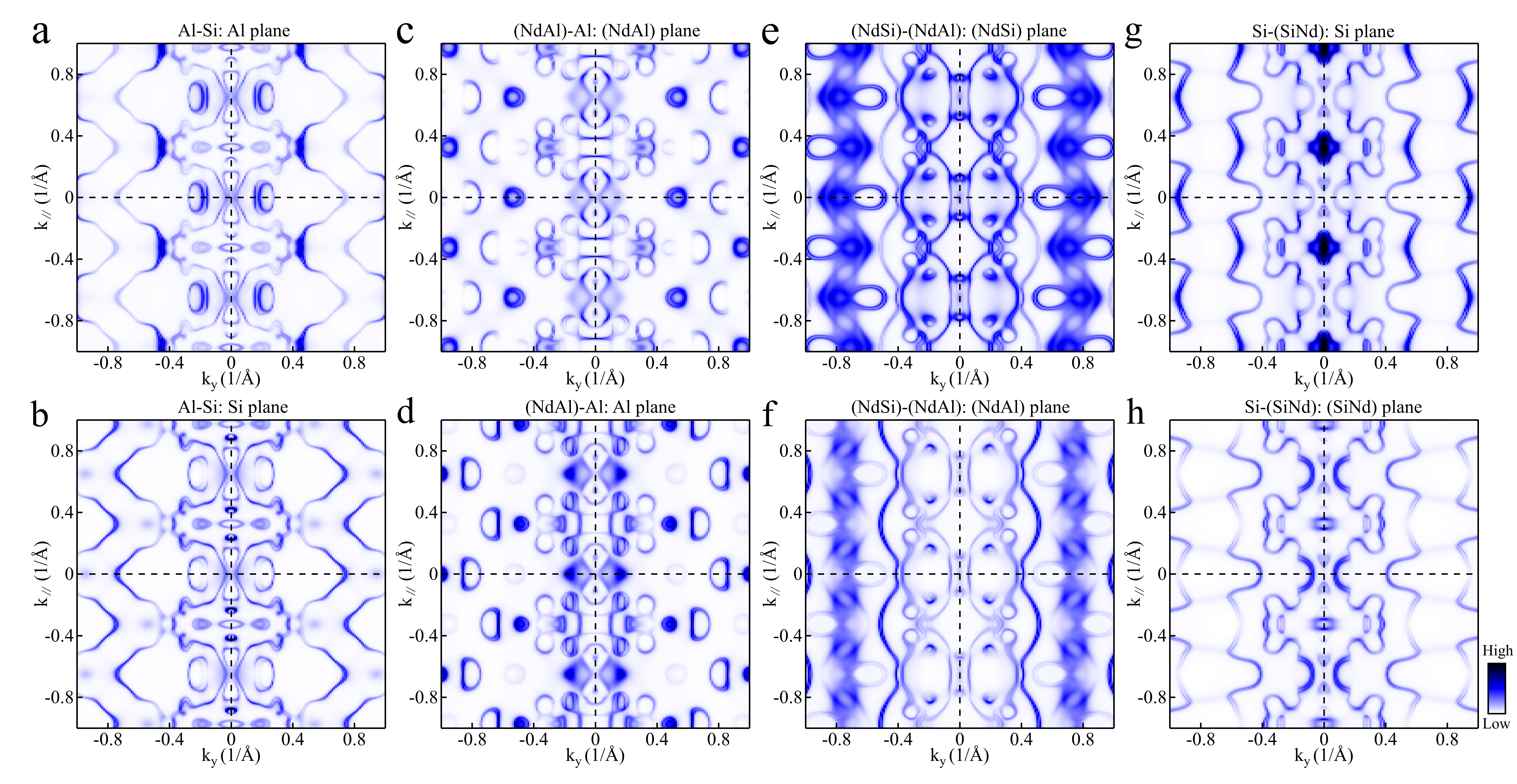}
\end{center}
\caption{Fermi surfaces calculated using a six-layer slab for the same cleavage terminations as in Fig.~\ref{S11}.
}
\label{S12}
\end{figure*}

\begin{figure*}[tbp]
\begin{center}
\includegraphics[width=1\columnwidth,angle=0]{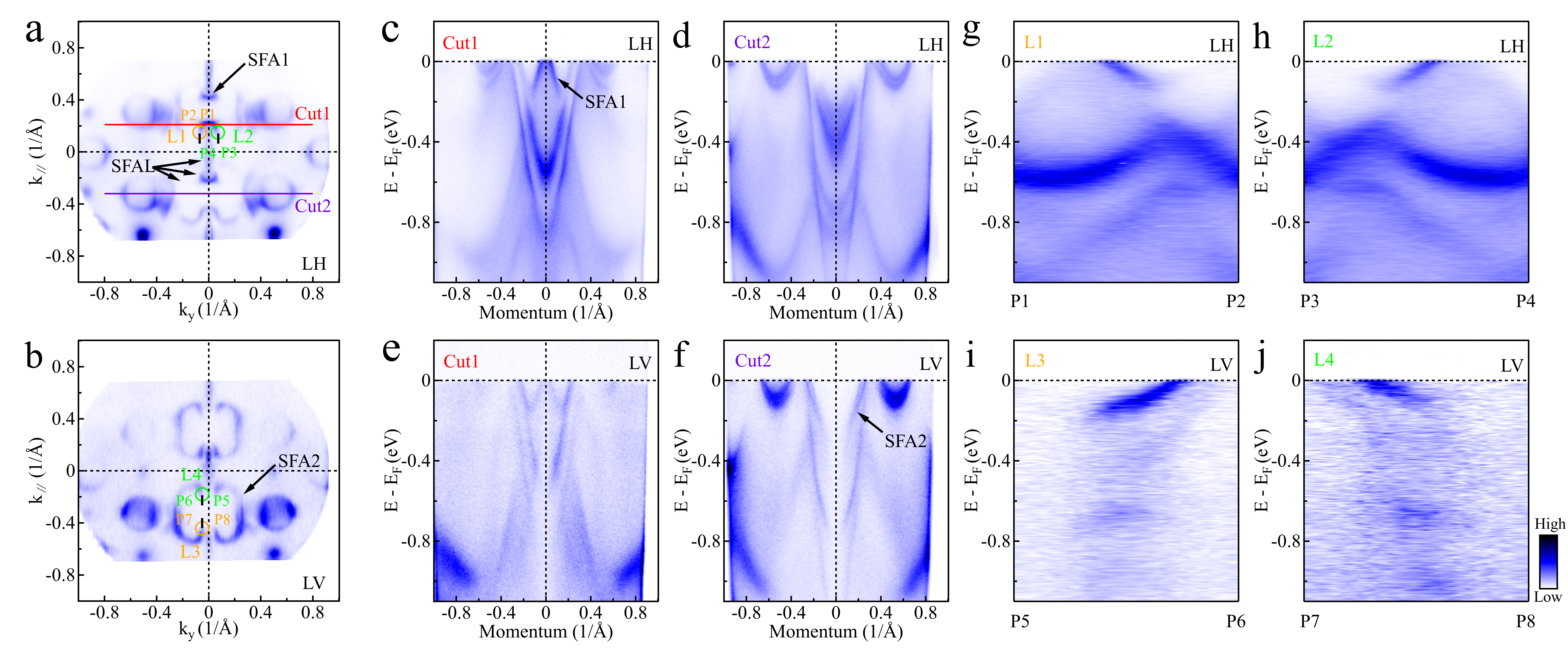}
\end{center}
\caption{\textbf{Identification of surface Fermi arcs.} (a-b) Fermi surfaces measured with LH (a) and LV (b) polarizations with defined momentum cuts. (c–f) Cuts through arc-like featuresreveal dispersions consistent with surface Fermi arcs (SFAs). (g-j) Cuts forming closed loops indicate nonzero Chern number and projected bulk charge, consistent with topological origin. On the low-symmetry (103) surface, hybridization between arcs from successive BZs can form closed SFALs marked by black arrows in (a).
}
\label{S13}
\end{figure*}

\begin{figure}[tbp]
\centering
\includegraphics[width=1\textwidth,angle=0]{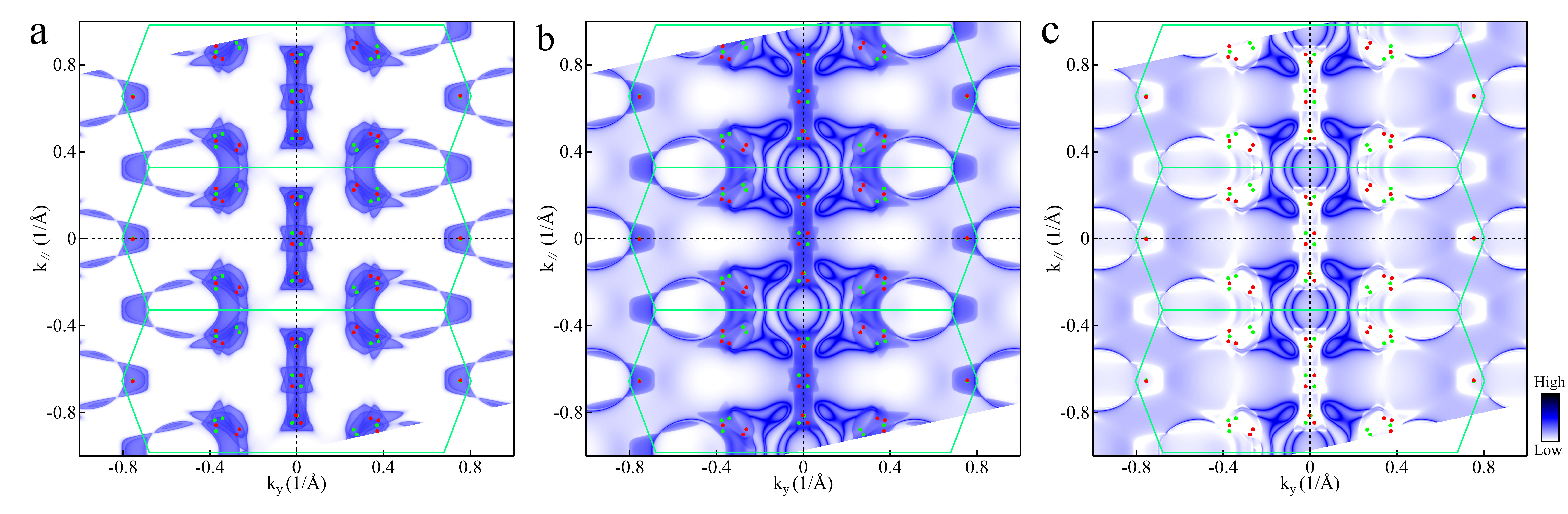}
\caption{\textbf{Calculated surface spectral functions and projected Weyl points on the (103) surface.} (a) SBPSs obtained from surface Green's function calculations. (b) Total surface spectral function including both SBPSs and true SSs. (c) Pure SSs obtained by subtracting the SBPS contribution in (a) from the total spectrum in (b). Red and green dots indicate the projections of Weyl points from the first-order to third-order bulk BZs onto the (103) SBZ, with red (green) corresponding to chirality $+1$ ($-1$). The green lines represent the SBZs. 
}
\label{S14}
\end{figure}

On the low-symmetry (103) surface, however, the situation becomes more subtle. Because successive bulk BZs project onto incommensurate positions, arcs from different zones can spatially overlap and hybridize. Such hybridization of surface Fermi arcs can occur in two ways. First, Weyl points from different bulk BZs may project onto the same surface location, while the SFA paths connecting them do not coincide, thereby forming an SFAL. Second, Weyl points from different bulk BZs project onto distinct surface locations, yet the projected surface Fermi arcs connecting them intersect and reconstruct on the surface, giving rise to a closed SFAL. In both cases, projection-induced coupling transforms open arcs into SFALs, which are no longer tied one-to-one to individual pairs of Weyl nodes. At the present highest ARPES resolution available to us, SFA1 and SFA2 in Fig.~\ref{S13}a-\ref{S13}b are seen to hybridize into a closed SFAL, and in addition a slender, elliptical closed SFAL (Fig.~\ref{S13}a) is discernible near the center of the SBZ. Clearer SFAL features can also be observed in the LD and CD Fermi surfaces shown in Fig.~\ref{S3}. As a result, the strict equality between the Chern number enclosed by a surface loop and the net projected chiral charge is relaxed. The bulk topology itself remains intact, but its manifestation on the surface is redistributed across an ensemble of hybridized states, including both open arcs and loops.\\

\textbf{Generalized BBC on low-symmetry facets} 

This motivates a more general understanding of BBC: instead of a direct correspondence between individual arcs and Weyl-point pairs, the correspondence must be understood collectively. The bulk Chern numbers remain faithfully encoded in the SSs, but their representation may take the form of hybridized, loop-like structures. This picture is consistent with the multi-BZ projection framework developed in Section 4, where higher-order replicas generate a superlattice periodicity commensurate with the SBZ, thereby restoring BBC in a broadened form.\\

\textbf{Physical implications of SFALs}

The emergence of SFALs enriches the boundary topology of Weyl semimetals and opens new physical possibilities. Closed momentum-space loops act as natural $k$-space cyclotron paths, giving rise to Shubnikov-de Haas and de Haas-van Alphen oscillations, with frequencies set by loop areas and angular dependence reflecting low-symmetry geometry. Magnetic breakdown between neighboring loops or between loops and arcs may generate composite oscillatory signatures. Overlapping loops also support phase-coherent interference, including Aharonov-Bohm-like oscillations in mesoscopic devices and interference patterns in pump-probe spectroscopy. Their extended and nearly parallel segments enhance nesting, favoring density-wave or pairing instabilities.

Moreover, reduced crystalline symmetry ensures that Berry-curvature contributions do not cancel, enabling finite Berry-curvature dipoles that can drive nonlinear Hall, rectification, and nonreciprocal transport responses. In proximity to superconductors, SFALs may mediate loop-selective Andreev reflection and unconventional Josephson interference distinct from open-arc transport. In STM quasiparticle interference, they should appear as sharp scattering vectors resembling conventional Fermi pockets, but decorated with spin-momentum locking. In nonequilibrium regimes, ultrafast ARPES could directly track their formation, hybridization, and recombination, offering insights into anisotropic relaxation and electron-phonon coupling.\\

{\bf 7. Mismatched periodicities and emergent superlattices}

The projection-induced formation of SFALs on low-symmetry facets can be understood within a broader context of emergent periodicities arising from mismatched lattices. A well-known analogue occurs in moir$\mathrm{\acute{e}}$ physics of stacked two-dimensional (2D) materials, where slight lattice mismatches or relative rotations generate emergent moir$\mathrm{\acute{e}}$ superlattices that fundamentally reshape the electronic structure. These moir$\mathrm{\acute{e}}$ patterns illustrate how new periodicities emerge from incommensurate or nearly commensurate structures, giving rise to flat bands, correlated phases, and unconventional topological responses.  

In crystalline Weyl semimetals, low-symmetry facets provide a closely related scenario: the mismatch between surface and bulk periodicities leads to apparent inconsistencies in BBC. These inconsistencies are resolved once multiple bulk BZ projections are taken into account, whose superposition generates emergent superlattices commensurate with the surface. This least common multiple construction of reciprocal vectors offers a systematic way to restore periodicity and reconcile surface and bulk descriptions.  

From this viewpoint, our framework can be naturally interpreted as establishing a momentum-space moir$\mathrm{\acute{e}}$ pattern. Each successive bulk BZ projection introduces a lateral shift relative to the SBZ, analogous to a real-space lattice mismatch. The accumulation of these shifts produces a long-period modulation in momentum space---a “moir$\mathrm{\acute{e}}$ stripe” pattern---that governs how Fermi arcs overlap, hybridize, and evolve into closed loops. Thus, the apparent paradox of BBC on low-symmetry facets is resolved by recognizing that the boundary spectrum is shaped by this emergent momentum-space moir$\mathrm{\acute{e}}$ superlattice.

This framework naturally extends to quasicrystals, which can be viewed as the extreme case of incommensurability. In quasicrystals, the reciprocal vectors form an intrinsically incommensurate set, so that no finite number of bulk BZs suffices to restore commensurability with the surface. Instead, quasi-periodicity emerges only in the infinite projection limit, where the dense set of reciprocal vectors reconstructs the correct boundary spectrum. From this viewpoint, quasicrystals are the limiting case of our framework: high-symmetry surfaces correspond to perfectly commensurate projections, low-symmetry surfaces require multiple bulk BZs to restore commensurability, and quasicrystals represent the fully incommensurate limit where only infinite projection recovers quasi-periodicity.  

This unified perspective highlights a general principle: mismatched periodicities--whether in moir$\mathrm{\acute{e}}$ bilayers, momentum-space moir$\mathrm{\acute{e}}$ stripes on low-symmetry crystal facets, or quasicrystals--generate emergent superlattices that govern hybridization and boundary states. In all cases, the interplay of commensurability, projection, and boundary conditions produces new forms of electronic structure beyond those expected from the parent lattice. The framework developed here therefore provides a common language that connects moir$\mathrm{\acute{e}}$ materials, crystalline surfaces of arbitrary orientation, and quasicrystalline systems, offering broad opportunities to uncover hidden boundary phenomena and design exotic responses across diverse classes of quantum matter.\\ 

{\bf 8. Determination of surface-bulk resonance states}

From a spectroscopic perspective, surface-bulk projected states (SBPSs) arise from projecting the three-dimensional (3D) bulk bands of the BZ onto the two-dimensional (2D) SBZ. In surface-sensitive probes such as ARPES, this projection appears as the integrated spectral weight over the out-of-plane momentum,
\[
A_{\mathrm{proj}}(\mathbf{k}_\parallel,E) \;=\; 
\int \frac{dk_\perp}{2\pi}\, W(\mathbf{k}_\parallel,k_\perp)\,
A_{\mathrm{bulk}}(\mathbf{k}_\parallel,k_\perp,E),
\]
where $A_{\mathrm{bulk}}$ is the bulk spectral function and $W(\mathbf{k}_\parallel,k_\perp)$ encodes the weight of bulk Bloch states on the top surface layers. SBPSs therefore follow the periodicity of the SBZ, yet they are not new eigenstates but rather the continuum projection of BSs, and they appear in ARPES as diffuse spectral backgrounds.

By contrast, SSs are sharp eigenmodes localized near the surface that exist inside bulk-projected band gaps. BSs, on the other hand, are fully extended Bloch states with negligible surface weight, which dominate spectra on strongly disordered or rough surfaces. When a SS lies outside the SBPS continuum, it remains sharp and well defined. However, when a SS enters the bulk-projected region, it hybridizes with the underlying bulk continuum and evolves into a surface-bulk resonance state (SBRS). SBRSs extend throughout the bulk but retain finite amplitude near the surface, enabling them to couple to surface-sensitive probes. As a result, they appear in ARPES as boundary-related features with surface periodicity but broader linewidths compared to pure SSs.

The identification of SBRSs requires separating them from SSs, SBPSs, and BSs. To achieve this, we adopt a sequence of complementary strategies.

\begin{figure*}[tbp]
\begin{center}
\includegraphics[width=1\columnwidth,angle=0]{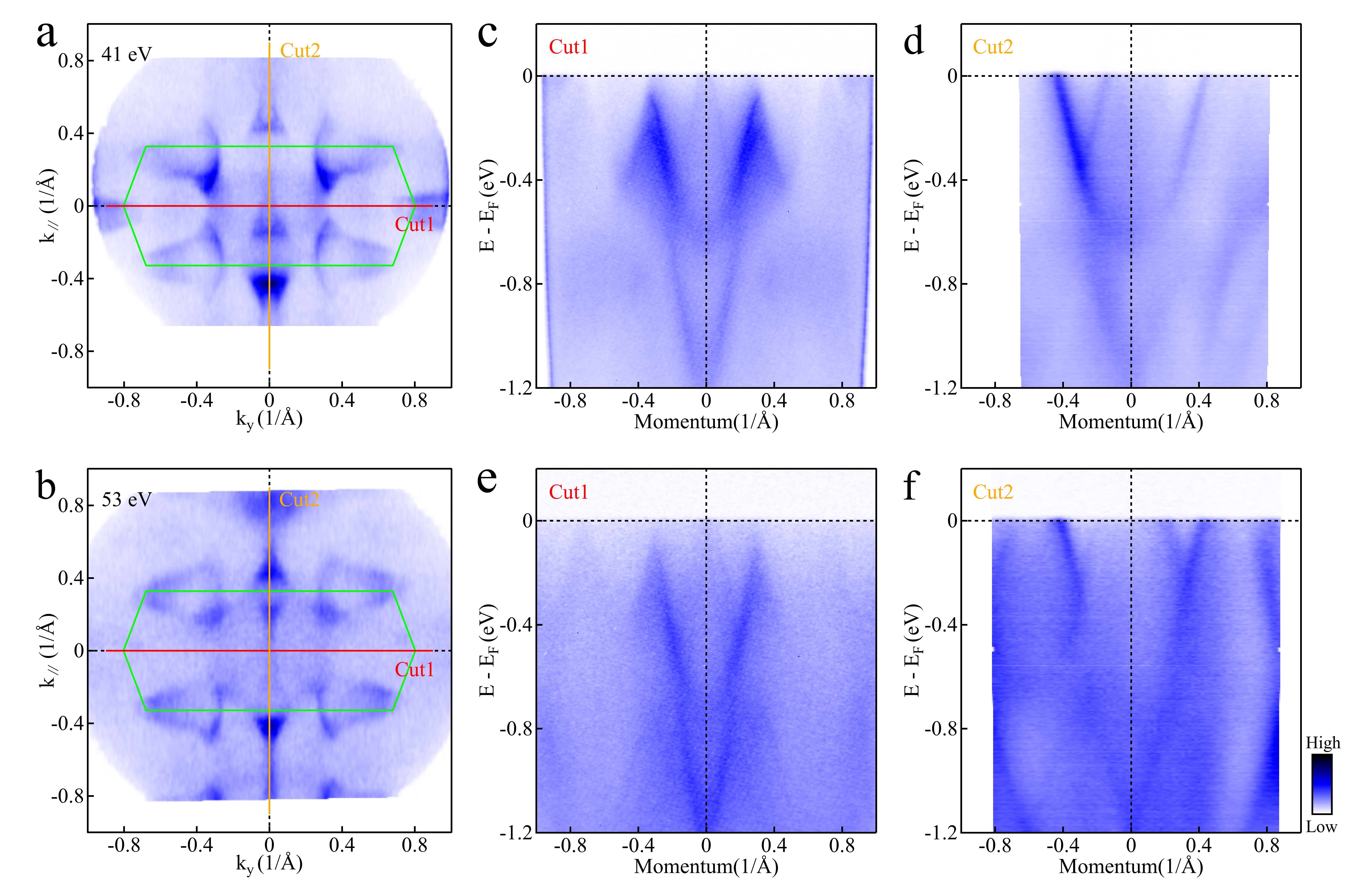}
\end{center}
\caption{\textbf{Suppression of SSs and emergence of surface-bulk resonance states.} (a-b) Fermi surface maps on well defined surfaces with photon energies of 41 eV (a) and 53 eV (b) after deliberate adsorption of impurities. (c–d) Corresponding band dispersions along high-symmetry cuts Cut1 (red) and Cut2 (orange) measured with photon energy of 41 eV. (e-f) Similar measurements as (c-d) but measured with photon energy of 53 eV.
}
\label{S15}
\end{figure*}

First, introducing surface disorder selectively suppresses coherent SSs while leaving SBPSs comparatively unaffected, since the latter originate from bulk projection. As shown in Fig.~\ref{S15}, deliberate adsorption of impurities onto initially flat surfaces strongly suppresses SSs, revealing the underlying SBPSs. Importantly, the diffuse SBPS features remain nearly unchanged between photon energies of 41 eV (Fig.~\ref{S15}a,~\ref{S15}c and~\ref{S15}d) and 53 eV (Fig.~\ref{S15}b,~\ref{S15}e and~\ref{S15}f), apart from variations in spectral weight caused by matrix-element effects or changes in the photon penetration depth, which alter the relative bulk versus near-surface contribution. This robustness further confirms the bulk-projected nature of SBPSs, in contrast to SSs that are highly sensitive to surface disorder.

\begin{figure*}[tbp]
\begin{center}
\includegraphics[width=1\columnwidth,angle=0]{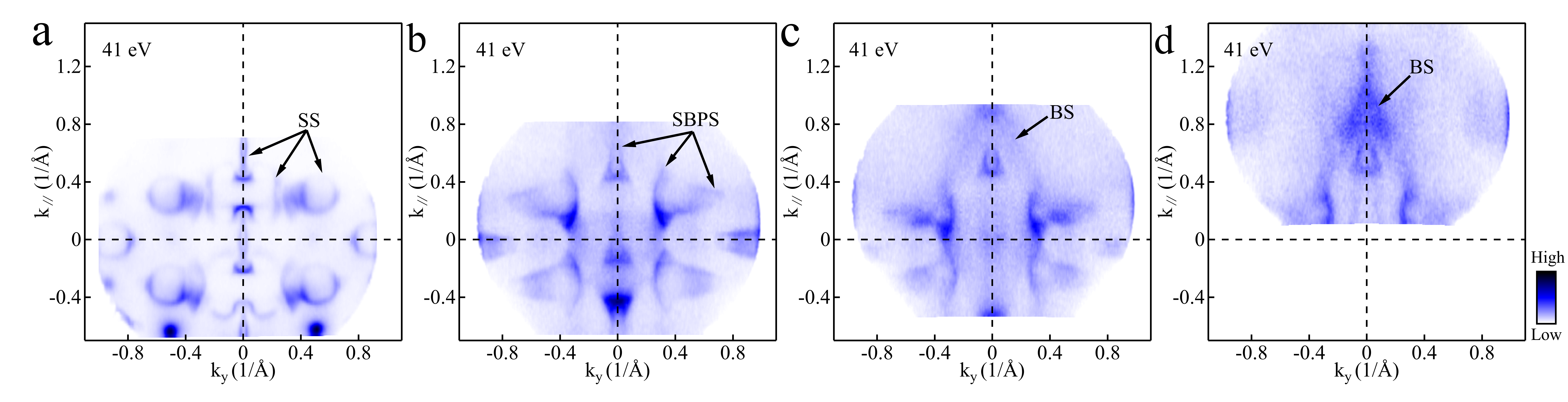}
\end{center}
\caption{Comparison of SSs, surface-bulk projected states (SBPSs), and bulk states (BSs) of NdAlSi (103) surface: (a) SSs dominated, (b) SBPSs dominated, (c-d) BSs dominated. All the Fermi surfaces measured with photon energy of 41 eV.
}
\label{S16}
\end{figure*}

Second, systematic comparison of different surface conditions distinguishes SSs, SBPSs, and BSs. As summarized in Fig.~\ref{S16}, (i) pristine flat surfaces are dominated by sharp SSs (Fig.~\ref{S16}a), (ii) moderately disordered surfaces expose SBPSs (Fig.~\ref{S16}b), and (iii) strongly disordered or rough surfaces suppress both SSs and SBPSs, leaving BSs dominant (Fig.~\ref{S16}c-\ref{S16}d). Notably, these comparisons also highlight a competition between SSs and SBPSs. On clean flat surfaces, coherent SSs dominate the spectral weight and partially obscure the diffuse SBPS background. When disorder is introduced, SSs are selectively suppressed, and the SBPS continuum becomes more visible. This redistribution of intensity indicates that SSs and SBPSs are not independent contributions but strongly intertwined: SSs emerge as coherent boundary modes in the absence of strong bulk projection, SBPSs dominate when SSs are weakened, and at their overlap the competition evolves into hybridization, which naturally leads to SBRSs (demonstrated below in Fig.~\ref{S18}).

\begin{figure*}[tbp]
\begin{center}
\includegraphics[width=1\columnwidth,angle=0]{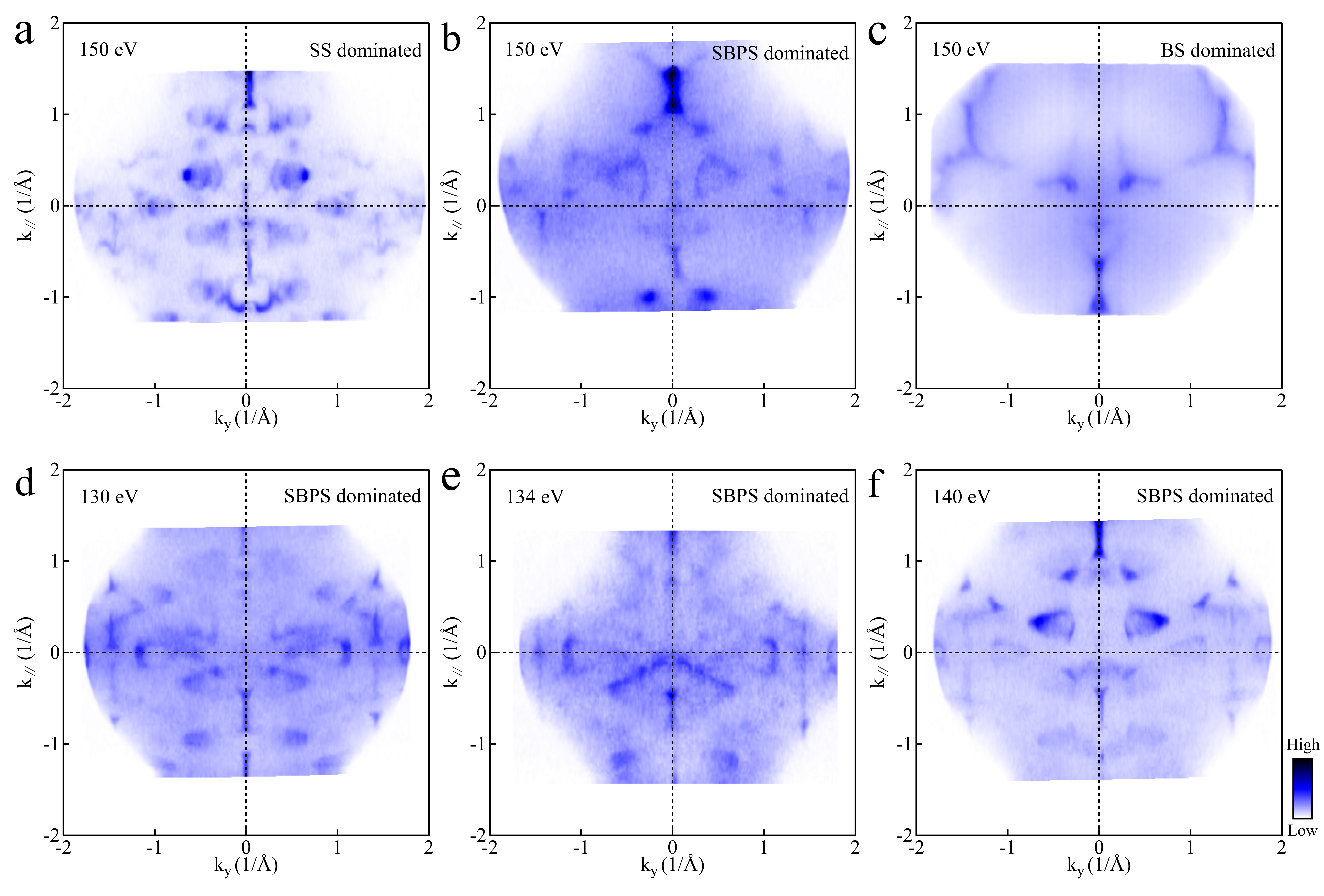}
\end{center}
\caption{\textbf{Photon energy dependence of Fermi surfaces highlighting SS, BS, SBPS contributions.} (a–c) Fermi surfaces measured at $h\nu=150$ eV under different surface conditions. (a) On a fresh and flat (103) surface, the spectrum is dominated by sharp SSs. (b) After mild impurity adsorption that introduces surface disorder, coherent SSs are suppressed, allowing SBPSs to dominate. (c) On an uneven surface region, the spectrum is instead dominated by incoherent BSs. (d–f) Fermi surfaces measured at $h\nu=130$ eV, 134 eV, and 140 eV, respectively, on a disordered flat surface. In all cases, SBPSs dominate the low-energy spectral weight, confirming their robustness across a wide photon-energy range.
}
\label{S17}
\end{figure*}

Third, photon energy dependence provides an additional check. As shown in Fig.~\ref{S17}, clean flat surfaces again display SS-dominated spectra (Fig.~\ref{S17}a), whereas mild disorder reveals SBPS features (Fig.~\ref{S17}b). On rough surfaces, BSs dominate instead (Fig.~\ref{S17}c). At intermediate photon energies of 130 eV-140 eV, disordered surfaces consistently show SBPS-dominated spectra (Figs.~\ref{S17}d–\ref{S17}f). These results confirm that SBPSs are robust projected continua, distinct from both SSs and BSs.

Finally, SBRSs are identified in Fig.~\ref{S18}, where photon energy dependent Fermi surface maps (41-53 eV) (Fig.~\ref{S18}a-\ref{S18}d) and corresponding dispersions (Fig.~\ref{S18}e-\ref{S18}l) along Cut1 and Cut2 are presented. When SSs overlap with the SBPS continuum, hybrid SBRS features emerge (Fig.~\ref{S18}a-\ref{S18}d), highlighted by their persistence across photon energies. The fact that SBRS dispersions remain nearly invariant with varying photon energies reinforces their identification as hybrid surface-bulk excitations, distinct from both SSs and pure BSs. 

Taken together, the complementary evidence from Figs.~\ref{S15}-\ref{S18} provides a clear and reproducible methodology for disentangling different classes of states. SSs are revealed on pristine surfaces, SBPSs appear as diffuse continua when SSs are suppressed, BSs dominate under strong disorder, and SBRSs emerge when SSs hybridize with the SBPS continuum. These results demonstrate that SBRSs form a distinct and intrinsic class of hybrid boundary excitations in NdAlSi and related Weyl semimetals.

Notably, the presence of pronounced SBRSs in NdAlSi may also explain why surface disorder is particularly effective in suppressing not only trivial SSs but also topological SSs (TSSs)\cite{CLi_PNAS2025_OTjernberg}. Although TSSs are protected by bulk band topology, in NdAlSi they frequently overlap with bulk-projected continua and thus evolve into SBRSs. In this hybridized form, their surface localization is weakened, rendering them highly susceptible to surface disorder that selectively suppresses the surface component. As a result, the apparent fragility of TSSs in ARPES is not inconsistent with their topological origin but rather reflects their strong hybridization with bulk continua through SBRS formation.\\

\vspace{3mm}

\newpage

\begin{figure*}[tbp]
\begin{center}
\includegraphics[width=1\columnwidth,angle=0]{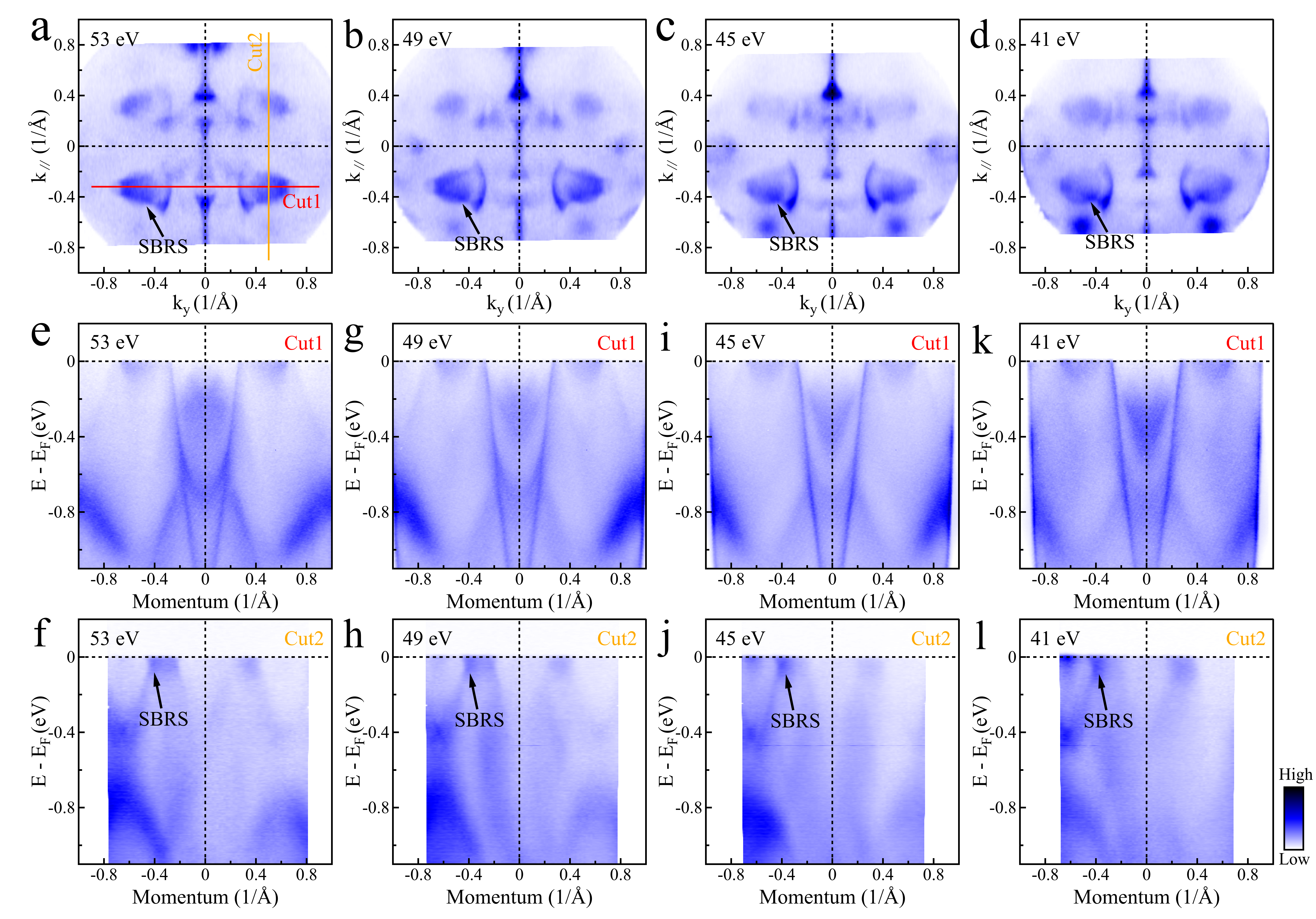}
\end{center}
\caption{\textbf{Distinguish of SBRS.} (a-d) Fermi surface maps recorded with photon energies of 53, 49, 45, and 41 eV, respectively. The sharp circular contours are attributed to SSs, while the broader triangular features originate from the SBPS. When the SSs overlap in energy and momentum with the SBPS continuum, they hybridize with the bulk Bloch states and evolve into SBRSs, as highlighted by the arrows. (e-l) The corresponding band dispersions taken along the momentum cuts Cut1 [red lines in (a)] (e, g, i, k)and Cut2 [orange lines in (a)] (f, h, j, l) defined in (a). 
}
\label{S18}
\end{figure*}